\documentclass[twocolumn,showpacs,aps]{revtex4}

\usepackage[dvips]{graphicx}

\usepackage{times}
\usepackage{color}
\begin{document}

\title{Optimal quantum parameter estimation in a pulsed quantum
optomechanical system}
\author{ Qiang Zheng $^{1,~2}$, Yao Yao $^{1,~3}$, and Yong Li $^{1,~4}$ }

\address{
$^{1}$ Quantum Physics and Quantum Information Division,
Beijing Computational Science Research Center, Beijing 100084, China
\\
$^{2}$ School of Mathematics and Computer Science, Guizhou Normal
University, Guiyang, 550001, China
\\
$^{3}$ Microsystems and Terahertz Research Center, China Academy of Engineering Physics, Chengdu Sichuan 610200, China
\\
$^{4}$ Synergetic Innovation Center of Quantum Information and Quantum Physics, University of Science and Technology of China,
Hefei, Anhui 230026, China
}

\begin{abstract}
We propose that a pulsed quantum optomechanical system can be applied for
the problem of quantum parameter estimation, which targets to yield higher
precision of parameter estimation utilizing quantum resource than that using
classical methods. Mainly concentrating on the quantum Fisher information
with respect to the mechanical frequency, we find that the corresponding
precision of parameter estimation on the mechanical frequency can be
enhanced by applying applicable optical resonant pulsed driving on the
cavity of the optomechanical system. Further investigation shows that the
mechanical squeezing resulting from the optical pulsed driving is the
quantum resource used in optimal quantum estimation on the frequency.
\end{abstract}

\maketitle

%\begin{CJK*}{GBK}{song}

%\thanks{Supported by the National Natural Science Foundation of
%China under Grant Nos. 11065005 and 11365006.
%\\*** Email: qzhengnju@gmail.com
% Tel:152-8601-8266; 182-8619-3565;
% Email: qz@gznu.edu.cn}

%\date{today}

\section{Introduction}

Quantum metrology~\cite{VGio11} is an active research field in recent years.
According to the quantum Cram\'{e}r-Rao inequality, the quantum Fisher
information (QFI) plays a key role in this subject \cite{Smerzi09, xmLu10,
yy14b,jin13, qz15}, which bounds the minimal variance of the unbiased
estimator. The QFI gives the quantum limit to the accuracy of the estimated
parameter with any positive-operator-valued-measure measurement. One of the
central ideas of quantum metrology is to beat the shot-noise limit and
approach the Heisenberg limit by virtue of quantum resource, such as quantum
entanglement or squeezing. There have been many studies on precision of
parameter estimation with sub-shot-noise limit in different physical
systems, such as the optical interferometers~\cite{Nagata07, Afek10, zyou14}%
, Bose-Einstein condensates~\cite{Strobel14}, atomic interferometers~\cite%
{wdli14}, and solid-state systems (e.g., the nitrogen-vacancy centres)~\cite%
{XYpan15, zhao13}. To the best of our knowledge, only a few papers~\cite%
{kiwa13, szang13, mtsan13} have devoted to investigating the quantum
metrology in the newly-developed novel quantum optomechanical device .%
%of optomechanical system.
%displayed that the cavity optomechanics, as an novel quantum device, can also be used to quantum metrology.

With the rapid advance of technology, quantum cavity optomechanics~\cite%
{tjkip08, pmey13, matjkfm14}, in which the mechanical resonator is coupled
to the optical field by radiation pressure or photothermal force, has
excited a burst of interest~\cite{masp14} due to the following two reasons:
On one hand, the cavity optomechanical system provides a new platform to
investigate the fundamental questions on the quantum behavior of macroscopic
system~\cite{lfbuch13} and even the quantum-to-classical transition~\cite%
{wmar03, cpsun07}; On the other hand, it brings a novel quantum device for
applications in ultra-high precision measurement~\cite{DRug04, careg08,
ganet09, sfors12, jmtay14}, gravitation-wave detection~\cite{aarv13},
quantum information processing~\cite{sman03} and quantum illumination \cite%
{shbar15}. Many interesting researches in cavity optomechanical systems,
such as optomechanically induced transparency~\cite{sweis10, gsash10},
ground-state cooling of the mechanical resonator~\cite{iwil07, cgenes08,
jdteufel10, jctpm11, yli14}, optomechanical entanglement~\cite{mpater07,
ltydw13}, optimal state estimation ~\cite{wwiesghof15}, have been reported.
These studies mainly rely on the enhanced coupling strength between the
phonic and photic fields by strongly pumping the optical cavity with a
continuous wave (CW) laser.

Different from the above case of CW laser driving, the so-called pulsed
quantum optomechanics~\cite{vanner11}, is also realized by driving the
optical cavity with (very) short optical pulses. Originally, this strategy
has been proposed in the systems of qubits~\cite{sl98}, and lately extended
to atomic ensembles~\cite{khammer05} and levitated microspheres trapped in
an optical cavity~\cite{oromeris11}. Compared with the CW-laser-driving
case, the benefit of the pulsed scheme is that it does not need the
existence of a stable steady-state for the optomechanical system. The pulsed
interaction has also displayed its superiority in preparation and
reconstruction of quantum state of the mechanical resonator~\cite%
{masj_aga14, jqliao11}, enhancing the optomechaical entanglement~\cite%
{hofer11, tapal13} and EPR steering~\cite{qyhe13}, and cooling the
mechanical mode~\cite{vanner13, smach12}.

Inspired by the experimental progress in pulsed quantum optomechanical
systems~\cite{vanner11, vanner13, tapal13},
%it's a natural idea to apply the pulsed quantum optomechanics to quantum parameter estimation.
it is a natural idea to investigate the high precision of parameter
estimation by applying applicable optical pulsed driving on the cavity of
the optomechanical system. Here we investigate a special pulsed
optomechanical system, where the coupling between the mechanical mode and
cavity field is quadratical to the mechanical motion and the cavity field is
resonantly driven with external optical pulses. We mainly focus on the QFI
with respect to the mechanical frequency, which is equivalent to estimating
on the mass of the mechanical resonator and could be used for mass precision
detection. With the Cram\'{e}r-Rao inequality, a larger QFI implies that the
mechanical frequency can be estimated with a higher precision. We show that
the QFI can be greatly enhanced when the period of the driving pulse matches
that of the mechanical motion. We also show that the mechanical squeezing
resulting from the resonant driving pulses is the quantum resource
strengthening the QFI.

This paper is organized as follows. The pulsed quantum optomechanical model
is discussed in Sec.~II. We investigate in Sec.~III the QFI of the pulsed
quantum optomechanical system with respect to the mechanical frequency.
Then, we display that the quantum squeezing is the resource used in optimal
quantum estimation. Finally, a summary is given in the last section. The
basic properties of the QFI, especially the QFI of a single-mode Gaussian
state, are reviewed in Appendix A.

\section{The pulsed quantum optomechanics}

%Although many physical implementations, such as suspended mirror in the
Optomechanical systems have been implemented in many physical systems, such
as suspended mirrors in the Fabry-P$\acute{e}$rot resonators~\cite{oarc06},
toroidal whispering gallery mode resonators~\cite{gantk09}, trapped
levitating nanoparticles~\cite{kies13}, ultracold atomic clouds in cavities~%
\cite{tppurdy10}. %, have been used to realize the optomechanics system,
Here we focus on a membrane-in-the-middle cavity optomechanical setup~\cite%
{jcsan10}, which has been used for quantum nondemolition measurement of the
phonon number state~\cite{jdthom08}, cooling of mechanical resonator~\cite%
{dengli08} or investigation of Landau-Zener-St$\ddot{u}$ckelberg dynamics~%
\cite{ghein10}. The linear and quadratical optomechanical couplings between
the cavity mode and the mechanical resonator can exist in this
membrane-in-the-middle optomechanical system.
Very recently, the optomechanical quadratical coupling is also
achieved in a crystal optomechanical system \cite{tkpmk15}, except for the membrane-in-middle setup.
%In this paper, we specifically consider the membrane-in-the-middle quadratical coupling setup, as shown in Fig.~\ref{system}.

\begin{figure}[!tb]
\centering
\includegraphics[width=3.5in]{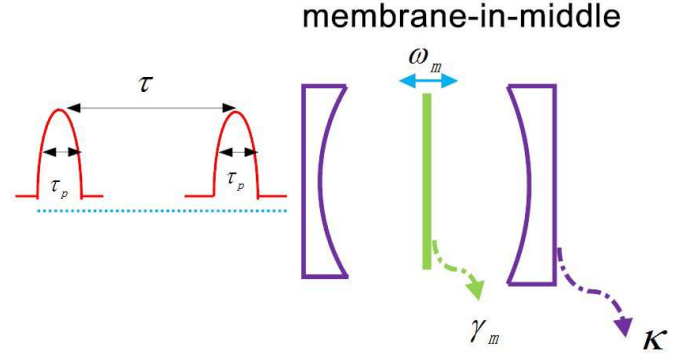} \hspace{2.5cm}
\caption{(Color online) Schematic diagram of the membrane-in-middle cavity
optomechanical setup considered in this paper. The coupling between the
cavity field (with the decay rate $\protect\kappa$) and the mechanical
resonator (with the resonance frequency $\protect\omega_{m}$ and the damping
rate $\protect\gamma_{m}$) is quadratical to the mechanical motion, and the
driving field is composed by a series of periodic pulses. The duration of
one pulse is $\protect\tau_{p}$, and the two consecutive pulses has the time
interval $\protect\tau$.}
\label{system}
\end{figure}

The membrane-in-the-middle quadratical coupling setup under consideration is
shown in Fig.~\ref{system} and the corresponding Hamiltonian is expressed as~%
\cite{pmeys08}
\begin{equation}
\begin{array}{l}
H=\frac{\hbar \omega _{m}}{2}(\hat{p}^{2}+\hat{q}^{2})+\hbar \omega _{c}\hat{%
a}^{\dag }\hat{a}+\hbar g_{2}\hat{a}^{\dag }\hat{a}\hat{q}^{2} \\
~~~~~+i\hbar \lbrack E_{0}(t)e^{-i\omega _{d}t}\hat{a}^{\dag }-h.c.].%
\end{array}
\label{ham}
\end{equation}%
Here $\omega _{m}$ is the frequency of the mechanical resonator, $\hat{p}$
and $\hat{q}$ are the dimensionless momentum and position operators
satisfying the relationship $[\hat{q},\hat{p}]=i$, %$i\hbar$
$\hat{a}$ is the annihilation operator of the cavity mode with resonance
frequency $\omega _{c}$ and decay rate $\kappa $, and $g_{2}$ is the
quadratic optomechanical coupling strength. Finally, $E_{0}(t)=\sqrt{%
2P_{0}(t)\kappa /(\hbar \omega _{c})}$ with $P_{0}(t)$
%being the time-dependent
the optical input power. We further assume that the cavity is driven
resonantly with $\omega _{d}=\omega _{c}$.

%%%%%%%%%%%%%%%%%%%%%%%%
%%%% Figure 1:
%%%%%%%%%%%%%%%%%%%%%%%%
%%%%%%%%%%%%%%%%%%%%%%%%
%%%%%%%%%%%%%%%%%%%%%%%%

For the mechanical resonator, by linearizing the optomechanical coupling,
the corresponding quantum Heisenberg-Langevin equation is obtained as
\begin{equation}
\begin{array}{llll}
\frac{d}{dt}\hat{q}=\omega _{m}\hat{p}, &  &  &  \\
\frac{d}{dt}\hat{p}=-\widetilde{\omega }_{m}(t)\hat{q}-\gamma _{m}\hat{p}%
+\xi , &  &  &
\end{array}
\label{Eqmir}
\end{equation}%
where $\widetilde{\omega }_{m}(t)=\omega _{m}+A(t)$ with $A(t)=2g_{2}n_{a}(t)
$ and $n_{a}(t)=\langle \hat{a}^{\dag }\hat{a}\rangle $. Here $\gamma _{m}$
is the mechanical damping rate, and $\xi $ denotes for the Browian noise
with null mean and correlation function%
%?¦Î(t)¦Î(t¡ä)?=((¦Ã_{m})/(2¦Ð¦Ø_{m}))¡Ò?^{¦¸_{c}}d¦Øe^{-i¦Ø(t-t¡ä)}¦Ø[coth(((?¦Ø)/(2k_{B}T)))+1] with ¦¸_{c} the cutoff frequency of the reservoir. This non-Markovian noise reduces to the a Markovian one <cite>garpol00</cite>
satisfying $\langle \xi (t)\xi (t^{\prime })\rangle =2n_{th}\gamma
_{m}\delta (t-t^{\prime })$ in the high-temperature limit $k_{B}T\gg \hbar
\omega _{m}$. Here $k_{B}$ is the Boltzmann constant, $T$ is the temperature
of the mechanical resonator, and $n_{th}=[e^{\hbar \omega
_{m}/(k_{B}T)}-1]^{-1}\approx k_{B}T/(\hbar \omega _{m})$ is the thermal
mean phonon number.

%From Eqs. (<ref>Eqmir</ref>), it's ready to obtain that the mean values of q and p equal to zero in the long-time limit for a fixed A. Linearizing Eqs. (<ref>Eqmir</ref>) near the null mean values,
From Eqs.~(\ref{Eqmir}), the dynamics of the second order moments of the
mechanical system
\begin{equation}
\begin{array}{llll}
\overrightarrow{v}(t)\equiv (\langle \hat{q}^{2}\rangle ,\langle \hat{p}\hat{%
q}+\hat{q}\hat{p}\rangle /2,\langle \hat{p}^{2}\rangle )^{\text{T}} &  &  &
\end{array}
\label{covmatm}
\end{equation}%
can be fully described by the equations
\begin{equation}
\begin{array}{llll}
\frac{d}{dt}\overrightarrow{v}(t)=\mathbf{U}_{A}\overrightarrow{v}(t)+%
\overrightarrow{N} &  &  &
\end{array}
\label{EqmirB}
\end{equation}%
for an initial Gaussian state of the mechanical resonator (such as the
thermal equilibrium state with mean phonon number $n_{th}$). Here the
superscript $\text{T}$ represents the transposition. Here
\begin{equation}
\mathbf{U}_{A}=\left(
\begin{array}{ccc}
0 & 2\omega _{m} & 0 \\
-\widetilde{\omega }_{m} & -\gamma _{m} & \omega _{m} \\
0 & -2\widetilde{\omega }_{m} & -2\gamma _{m}%
\end{array}%
\right) ,  \label{matrixU}
\end{equation}%
and $\overrightarrow{N}=(0,0,(2n_{th}+1)\gamma _{m})^{\text{T}}$.

The solution of Eq.~(\ref{EqmirB}) is formally expressed as
\begin{equation}
\begin{array}{llll}
\overrightarrow{v}(t)=e^{\mathbf{U}_{A}t}\overrightarrow{v}%
(0)+\int_{0}^{t}e^{\mathbf{U}_{A}(t-t^{\prime })}\overrightarrow{N}%
dt^{\prime } \\
~~~~~~~~~=\mathbf{M}_{A}(t)\overrightarrow{v}(0)+\overrightarrow{v}_{\text{%
inh}}. &  &  &
\end{array}
\label{noisv}
\end{equation}%
Here $\mathbf{M}_{A}(t)=e^{\mathbf{U}_{A}t}$ and $\overrightarrow{v}_{\text{%
inh}}=\mathbf{U}_{A}^{-1}[\mathbf{I}_{3}-\mathbf{M}_{A}(t)]\overrightarrow{N}
$ with $\mathbf{I}_{3}$ being $3\times 3$ unitary matrix.

Now, we study the case that the driving field is the periodic Gaussian
pulses with the duration $\tau _{p}$ and period $\tau $, i.e. $%
P(t)=P_{0}\sum_{n}\exp [-(t-n\tau )^{2}/\tau _{p}^{2}]$. Here we keep the
condition $1/\tau _{p}<c/2L$ ($L$ is the cavity length), which means the
optical driving pulses %with small bandwidth
will not excite the near cavity modes except the desired one and thus the
cavity field can be always considered as a single mode one. For remaining
the quadratic coupling during the pulsed driving, the membrane should be
locked at a cavity node. Accordingly the effective frequency of the
mechanical resonator is periodically modulated in time via the optical
driving pulses. An alternative scheme to achieve periodic modulation of the
effective frequency of the mechanical resonator is given in Ref.~\cite%
{ankb08} with a two-tone drive.

Moreover, we assume that the system works in the following parameter
regimes: (i)~$1/\tau \ll \kappa $, (ii)$~1/\tau _{p}\ll \kappa $,~(iii)~$%
\tau _{p}\ll 1/\omega _{m}$. The condition (i) implies the cavity is rapidly
excited by one pulse and damps to the vacuum state before the next pulse
arrives~\cite{masj_aga14}. And the condition (ii) means that the bandwidth
of the pulses is much smaller than that of the cavity, which guarantees the
pulse entering into the cavity spectrally. The last condition (iii) makes
sure that the free rotation of the mechanical resonator is frozen in the
process of the pulse-mirror interaction. In this case, the intracavity
photon number can be approximated as a series of Dirac delta functions $%
n_{a}(t)\propto \sum_{n=0}^{{}}\delta (t-n\tau )$ in the typical evolution
time of the mechanical resonator. As a result, the whole dynamics of the
optomechanical system is divided into two steps: (1) one kick at time $%
t=n\tau $, which can be described by the unitary operator $U_{\text{K}%
}=e^{-i\theta \hat{q}^{2}}$ with $\theta =g_{2}\int_{\Delta t}n_{a}(t)dt$
being the kick strength ($\Delta t$ means the integral time domain and is of
the order of the typical time for a Gaussian pulse) and thus corresponds to
the linear transformation $q\rightarrow q$ and $p\rightarrow p-2\theta q$,
and (2) the free-evolution lasting time $\tau $ between two adjacent kicks,
whose corresponding evolution is determined by Eq. (\ref{Eqmir}) with $A(t)=0
$.

Combined these two evolving processes, the equation of motion in a $\tau $
circle is given as \cite{masj_aga14}
\begin{equation}
\overrightarrow{v}((n+1)\tau )=\mathbf{M}_{0}(\tau )\mathbf{K}%
\overrightarrow{v}(n\tau )+\overrightarrow{v}_{\text{inh}}(\tau ),
\label{combkickU}
\end{equation}%
where $\mathbf{M}_{0}(\tau )\equiv \mathbf{M}_{A=0}(t)|_{t=\tau }$, and
\begin{equation}
\mathbf{K}=\left(
\begin{array}{ccc}
1 & 0 & 0 \\
-2\theta & 1 & 0 \\
4\theta ^{2} & -4\theta & 1%
\end{array}%
\right)  \label{matrixK}
\end{equation}%
denoting the effect of the kick on the second order moments. That is, $%
\mathbf{K}$ is the representation of $U_{\text{K}}$ based on the second
order moments. Making use of Eq.~(\ref{combkickU}), the stroboscopic state
of the mechanical resonator at time $t=n\tau $ is obtained as
\begin{equation}
\begin{array}{llll}
\overrightarrow{v}(n\tau )=(\mathbf{M}_{0}(\tau )\mathbf{K})^{n}%
\overrightarrow{v}(0) &  &  &  \\
~~~~+[\mathbf{I}_{3}-(\mathbf{M}_{0}(\tau )\mathbf{K})^{n}](\mathbf{I}_{3}-%
\mathbf{M}_{0}(\tau )\mathbf{K})^{-1}\overrightarrow{v}_{\text{inh}}(\tau ).
&  &  &
\end{array}
\label{solB}
\end{equation}

\section{QFI of the pulsed optomechanics}

After detailed presentation of the pulsed quantum optomechanical model in
the previous section, here we move to investigate the quantum parameter
estimation
%with respect to the mechanical frequency in this model, and concentrate on
via the related QFI in this model. We will also show that the quantum
resource used for parameter estimation is the squeezing produced by pulsed
driving.

\subsection{Primary discussions}

The parameter to be estimated in this paper is the frequency $\omega _{m}$
of the (harmonic) mechanical resonator. Choosing this parameter is based on
the following consideration. With the relation $\omega _{m}=\sqrt{k_{m}/M}$ (%
$k_{m}$ and $M$ being the spring constant and the mass, respectively), the
QFI with respect to $M$ is proportional to that of with respect to $\omega
_{m}$, that is
\begin{equation}
\begin{array}{llll}
F_{\text{M}}=\mu F_{\omega _{m}}, &  &  &
\end{array}
\label{fscalA}
\end{equation}%
where $\mu =\frac{k_{m}}{4M^{3}}$ is the scaling factor. As a result, the
estimation on $M$, just as done in the mass spectrometer~\cite{kdzhu13}, is
equivalent to the estimation on $\omega _{m}$. In principle, the parameter
to be estimated in this model can be the ones other than $\omega _{m}$ ($m$%
). Here we just concentrate on the case of $\omega _{m}$ via its related QFI
$F\equiv F_{\omega _{m}}$.

The numerical values of the parameters used in this paper are based on the
state-of-the-art experiments reported in Ref.~\cite{neflow12}. We choose the
cavity decay as $\kappa \simeq 10^{2}$\,GHz, and the driving pulses with
duration $\tau_{p}=0.1$\,ns. We also set the mechanical frequency $%
\omega_{m}=0.5 \times 10^{6}$\,Hz and the damping rate $\gamma_{m}=10^{2}$%
\,Hz unless otherwise stated. By carefully choosing the mechanical mass,
reflectivity, and initial equilibrium position, the kick strength $\theta$
is in the range of $(0.01, 10)$ for the typical coupling strength $g_{2}$.

\subsection{Numerical results of the QFI}

With substituting Eq.~(\ref{solB}) into Eq.~(\ref{fisherGau}), the QFI $F$
can be obtained straightforwardly. Fortunately, the last term in Eq.~(\ref%
{fisherGau}) vanishes since there is no first-order moment of the mechanical
motion. In what follows, we mainly explore $F$ by numerical simulations as
the analytical solution is too cumbersome.

\begin{figure}[!tb]
\centering
\includegraphics[width=2.5in]{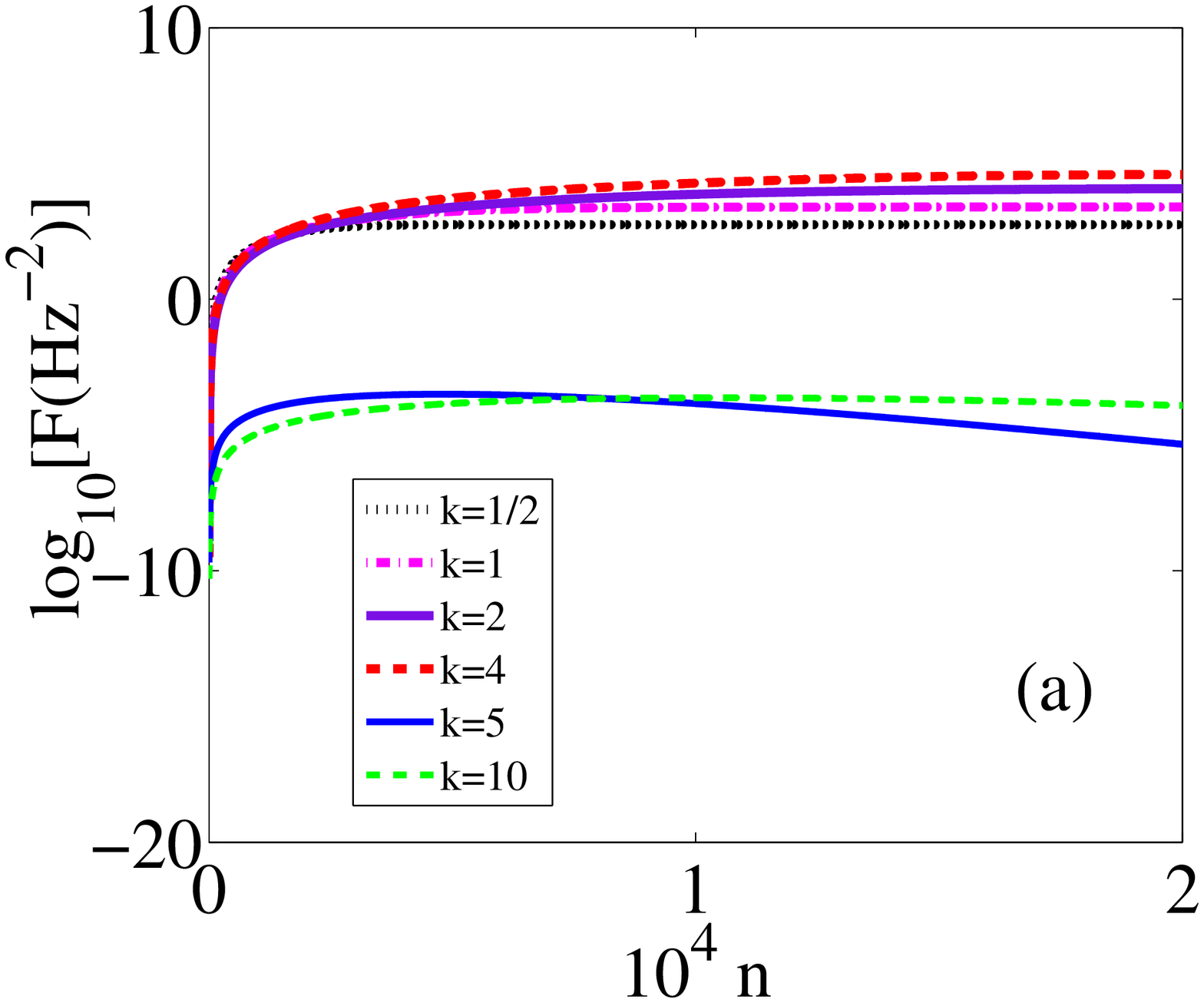} %
\includegraphics[width=2.5in]{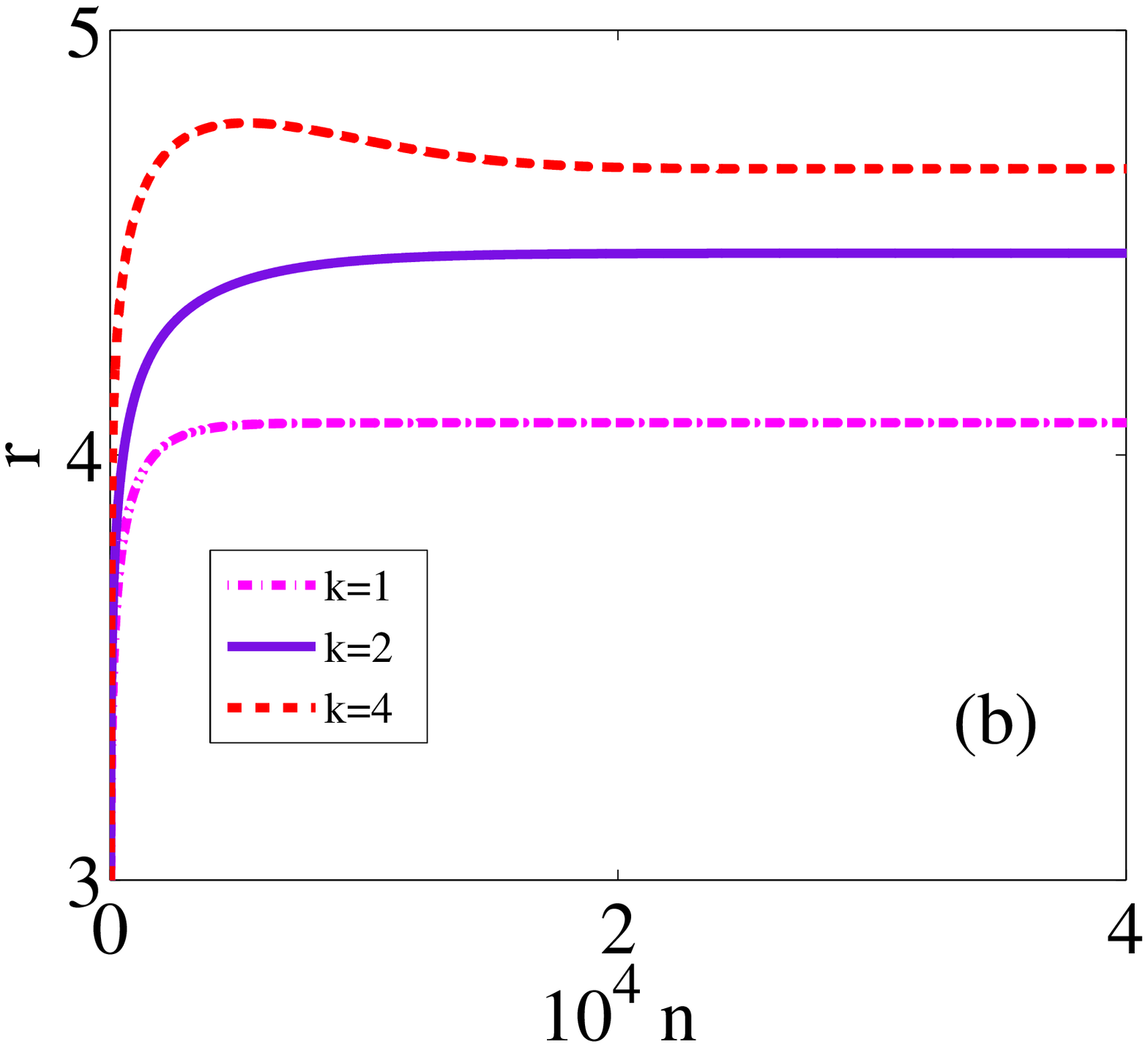}
\hspace{2.5cm}
\caption{(Color online) (a) and (b) corresponds to the variation of the QFI $%
F$ and the squeezing degree $r$ defined in Eq.~(\protect\ref{squstreng}) in
terms of the pulse number $n$, respectively. Different lines corresponds to
the different values of the parameter $k=T_{0}/\protect\tau$ ($\protect\tau$
being the period of the pulses and $T_{0}=2 \protect\pi/\protect\omega_{m}$%
), respectively. The other parameters are $n_{th}=100$ and $\protect\theta%
=1.0$.}
\label{fishfigA}
\end{figure}

With the extensive numerical simulations, we find that the evolution of $F$
in terms of the pulse number $n$ shows two distinct behaviors, as shown in
Fig.~\ref{fishfigA}(a). In this figure, the period of pulse $\tau$ matches
that of the mechanical resonator $T_{0}$ ($\equiv {2\pi}/{\omega_{m}})$
though $\tau={T_{0}}/{k}$, with $k$ taking the representative values $\{
\frac{1}{2},~1,~2,~4,~5,~10 \}$ denoting the pulse number in one period of
the mechanical motion. With $k\leq 4$, the QFI $F$ increases very quickly at
the initial pulse number $n$ and arrives to a large constant value in the
large $n$ limit. However, with the increase of $k\geq 5$, the QFI $F$ shows
the behaviors of initially increasing with $n$ and then gradually going down
to zero with large $n$. In this case, the value of the QFI is very small
compared to that of $k\leq 4$. When the pulse period doesn't match the
mechanical period, e.g., ${k}\equiv\frac{T_{0}}{\tau}$ is an irrational
number, we also find that the value of QFI $F$ becomes to much more smaller,
compared to the periods match case. More importantly, it is shown in Fig.~%
\ref{fishfigA}(a) that the QFI $F$ with $k=4$ is optimal, and the reason of
this optimal will be discussed in Set.~III B.

%%%% Figure 2:
%%%%%%%%%%%%%%%%%%%%%%%%
%%%%%%%%%%%%%%%%%%%%%%%%
%%%%%%%%%%%%%%%%%%%%%%%%

The physics of these behaviors of the QFI $F$ can be understood as
following. %First of all, notice
Note that the parameter to be estimated is the mechanical frequency $%
\omega_{m}$. If the period of the driving pulses $\tau$ matches that of the
mechanical period $T_{0}$, the information of $\omega_{m}$ can be extracted
to the greatest extent. As a result, the value of QFI $F$ is large
necessarily in the matching cases. A constant value of the QFI in the large $%
n$ limit origins from the balance between the pulsed driving and the
mechanical damping.

\begin{figure}[!tb]
\centering
\includegraphics[width=2.5in]{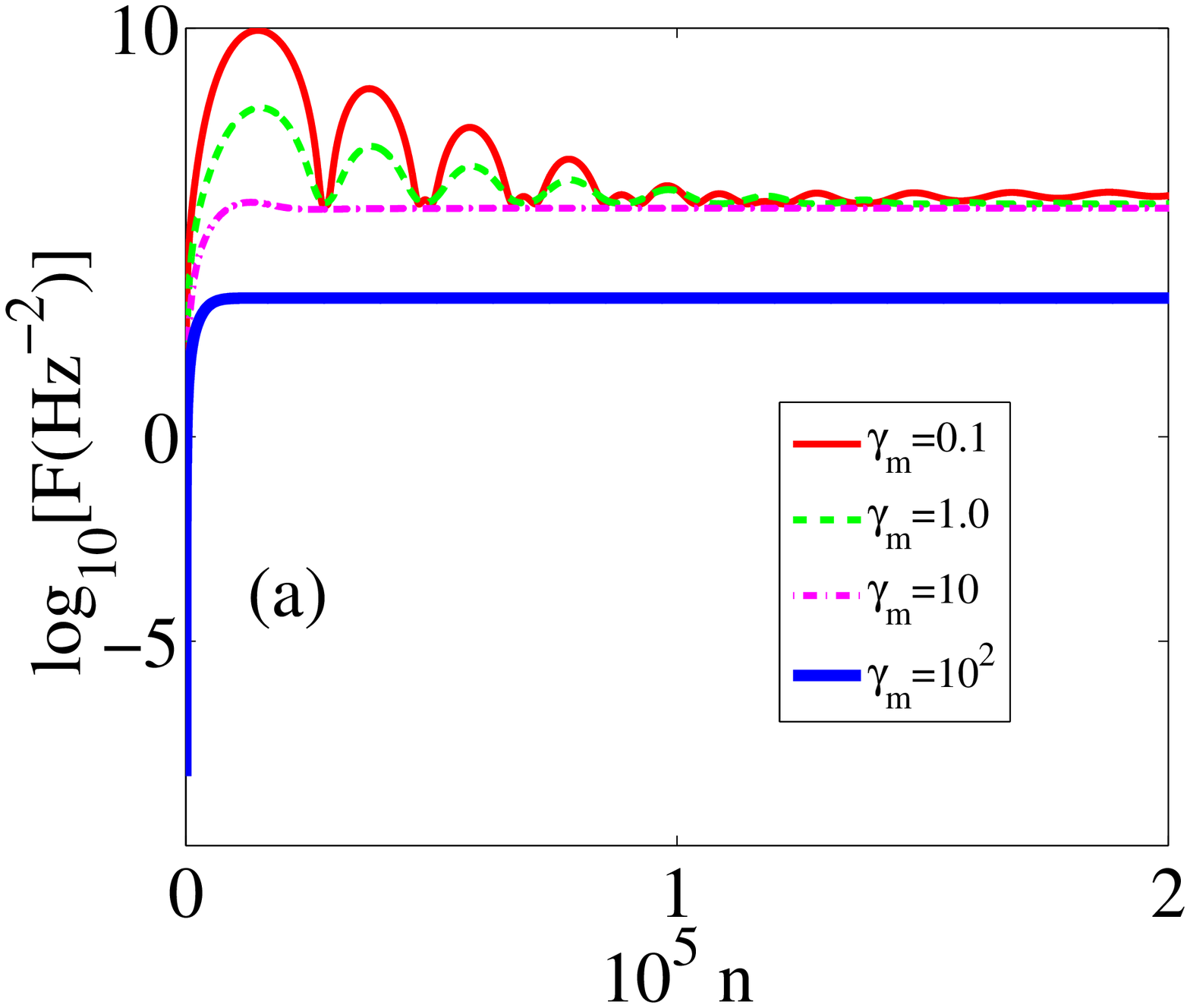} %
\includegraphics[width=2.5in]{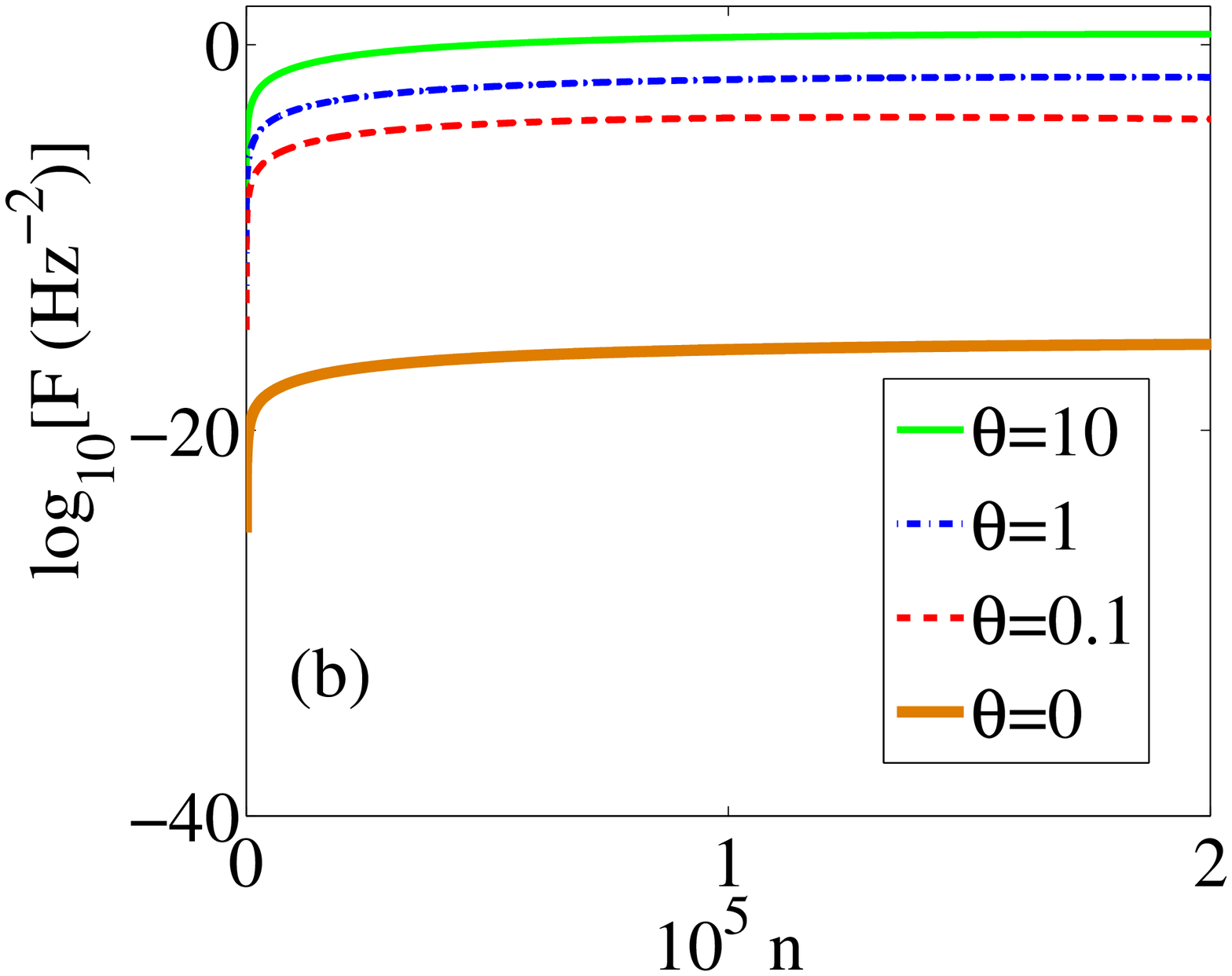} \hspace{3.5cm}
\caption{(Color online) The QFI $F$ as a function of the pulse number $n$
with different decay rate $\protect\gamma_{m}$ (in units of Hz) for $\protect%
\theta=1.0$ and $k\equiv T_0/\protect\tau =1$ (a) and with different kick
strength $\protect\theta$ for $\protect\gamma_{m}=10^{2}$\,Hz and $\protect%
\tau=10^{-7}$\,s (b). The other parameter is $n_{th}=100$. }
\label{fishfigD}
\end{figure}

The influence of the mechanical decay rate $\gamma_{m}$ on the QFI $F$ is
studied in the resonant driving regime, as shown in Fig.~\ref{fishfigD}(a).
This figure displays that with the increase of $\gamma_{m}$, the value of $F$
decreases considerably. Moreover, $F$ displays the oscillation behavior when
$\gamma_{m}$ is very small. The reason of this oscillation is simple: the
coherent evolution of the mechanical resonator, determined by $\omega_{m}$,
is dominated if the mechanics has very high quantity factor $\frac{\omega_{m}%
}{\gamma_{m}}$.

%%%% Figure 3:
%%%%%%%%%%%%%%%%%%%%%%%%
%%%%%%%%%%%%%%%%%%%%%%%%
%%%%%%%%%%%%%%%%%%%%%%%%

The effect of the kick strength $\theta$ on the QFI $F$ is shown in Fig.~\ref%
{fishfigD}(b). It is obvious that $F$ also goes down considerably with the
decrease of $\theta$. This can be easily understood: without external
driving, the mechanical damping will suppress its coherence in the long-time
limit. As a result, it is natural that the QFI $F$ decreases.

\begin{figure}[tb]
\centering
\includegraphics[width=2.5in]{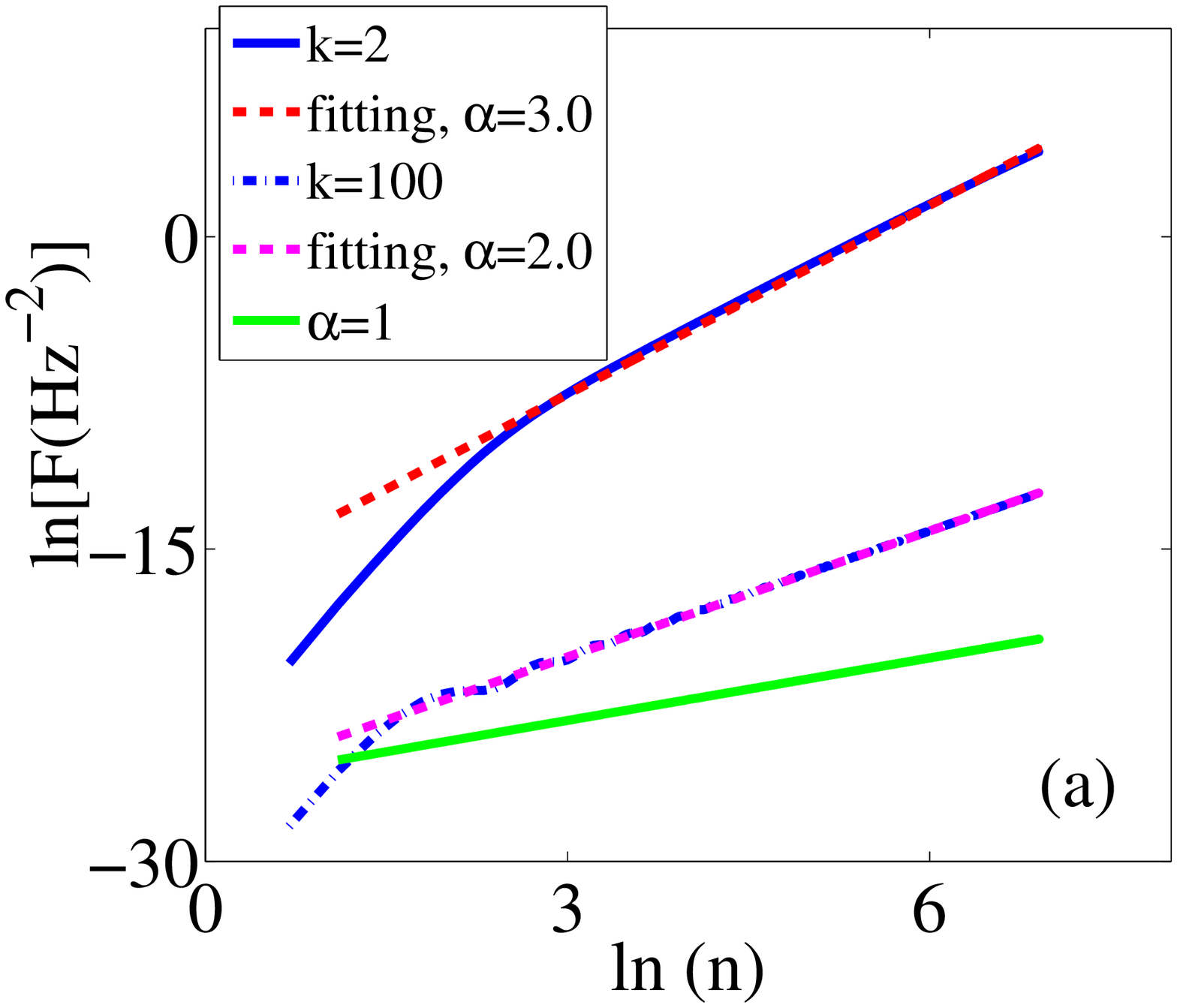} %
\includegraphics[width=2.5in]{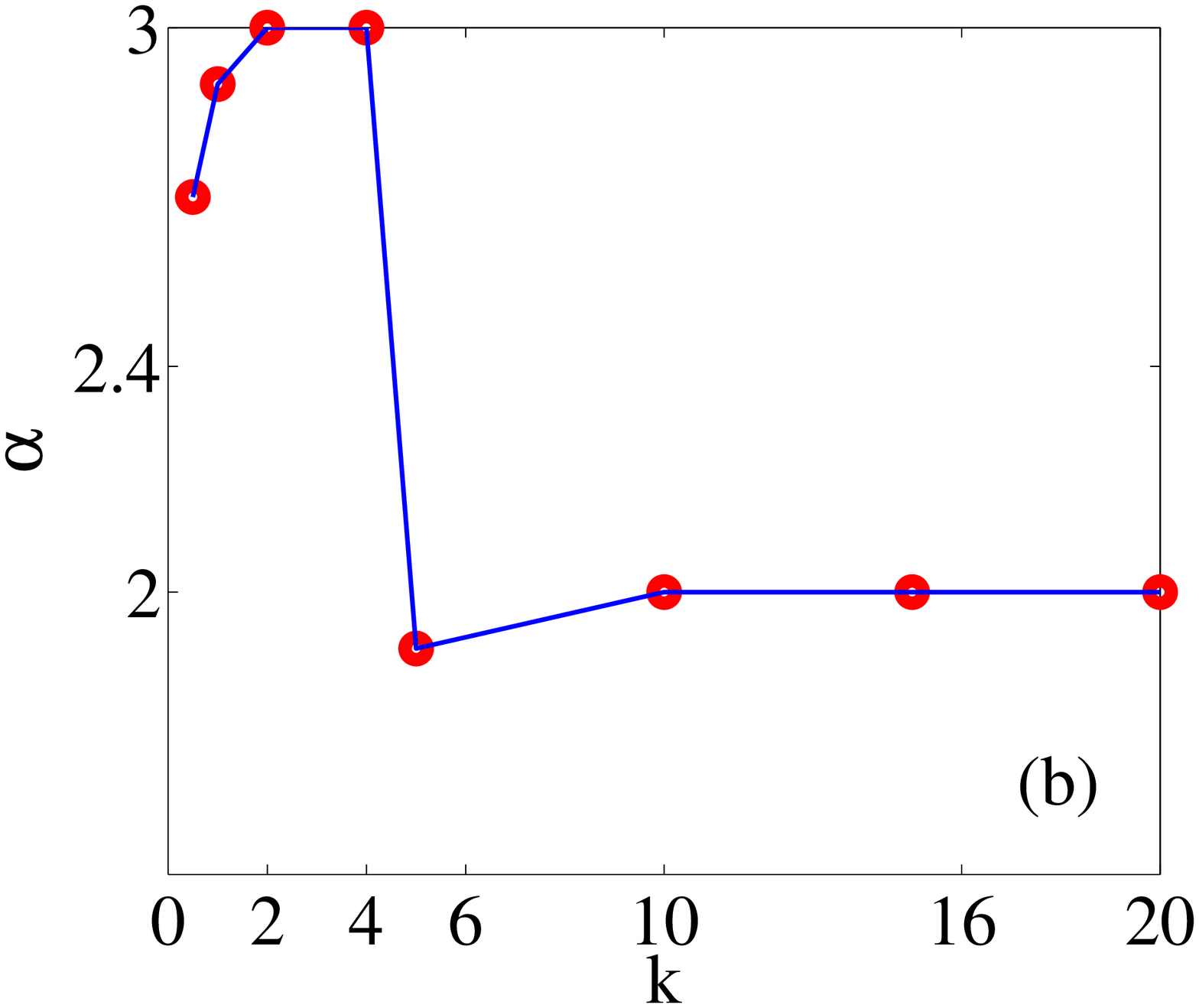} \hspace{3.5cm}
\caption{(Color online) (a) Fitting the growing-up part of
$F$ with respect to the pulse number $n$. The index $\protect\alpha $
is determined by numerically fitting $F\propto n^{\protect\alpha }$. Both
the x-axis and y-axis are scaled based on the natural logarithm. The
shot-noise limit ($\protect\alpha =1$) is displayed as the green line. (b)
The dependence of the index $\protect\alpha $ on the parameter $k$. The
other parameters are the same as that in Fig.~\protect\ref{fishfigA}. }
\label{fishfigC}
\end{figure}

In order to show the advantage of our pulsed-driving estimation protocol, we
study in Fig.~\ref{fishfigC}(a) the relationship between the growing-up part
of the QFI $F$ and the pulse number $n$. We find that $F\propto n^{\alpha }$
by numerically fitting, with the index $\alpha$ dependent on the parameter $k
$, as shown in Fig.~\ref{fishfigC}(b). Moreover, we also checked numerically
that this dependence of $\alpha $ on $k$ is independent of the parameters $%
\gamma _{m}$ and $\theta$.

From Fig.~\ref{fishfigC}(b), it's clear that the QFI $F$ with
respect to the pulse number $n$ approaches the Heisenberg limit with
$\alpha =2$ for the relatively large $k>10$, similar to the results
obtained previously in the systems of the pulsed driving qubit
\cite{Ueda10, tan13}. For $k=2$, $4$, the QFI shows the behaviour
beyond the Heisenberg limit \cite{sbad082} with $\alpha =3$. The
Heisenberg limit and beyond it display that the pulsed driving is an
essential way to enhance the precision of parameter estimation.

%%% Figure 4:
%%%%%%%%%%%%%%%%%%%%%%%%
%%%%%%%%%%%%%%%%%%%%%%%%
%%%%%%%%%%%%%%%%%%%%%%%%

In the following, we discuss how to read out the mechanical quadratures
experimentally, which are required for the QFI of the mechanical resonator.
For the pulsed optomechanical system, there exist at least two
experimentally feasible schemes to achieve this aim. The main idea of the
first one \cite{vanner13} is based on the homodyne detection of the mixing
between the signal pulses, which interact with the mechanical resonator, and
the local oscillator (LO) pulses, as shown in Ref.~\cite{vanner13} (wherein
Fig. 1(a)). The second scheme is based on the beam-splitter interaction,
where the mechanical quadratures are transferred into the optical readout
pulses (latter injected), as displayed in Ref.~\cite{tapal13} (wherein
Fig.~2). Thus, the information of the mechanical resonator can be gained
with the output of the readout pulses.

\section{Mechanical squeezing as a quantum resource}

Generally, it is well-known that the squeezed state~\cite{jmaxgw11}, as an
essential resource for quantum metrology~\cite{caves81,asmer13}, can enhance
the precision of parameter estimation. Motivated by this fact, we study the
relationship between the QFI $F$ and the mechanical squeezing in this
section.

Any single-mode Gaussian state can be expressed as~\cite{cweed12}
\begin{equation}
\begin{array}{llll}
\rho=\hat{D}(\alpha) \hat{S}(r, \phi) \rho_{th}(\overline{n}) \hat{S}%
^{\dag}(r, \phi) \hat{D}^{\dag}(\alpha), &  &  &
\end{array}%
\end{equation}
where $\hat{D}(\alpha)=\exp[\alpha \hat{a}-h.c.]$ is the displacement
operator of bosonic mode $\hat{a}$, $\hat{S}(r,\phi)=\exp[\frac{r}{2}(e^{-2i
\phi} \hat{a}^2-h.c. )]$ is the squeezing operator, and $\rho_{th}(\overline{%
m})=\sum_{m=0}^{\infty}\frac{\overline{m}^2}{( \overline{m}+1)^{m+1}} |m
\rangle \langle m|$ denotes the thermal state with $\overline{m}$ the mean
particle number. The squeezing strength $r$ and the squeezing angle $\phi$
are determined by
\begin{equation}
\begin{array}{llll}
r=\frac{1}{2} \mathrm{arcsinh} [\frac{1}{2} ( \frac{\gamma}{\det {%
\mbox{\boldmath{$\Sigma$}}}_{\varphi} })^{\frac{1}{2}} ], &  &  &
\end{array}
\label{squstreng}
\end{equation}
\begin{eqnarray}
2\phi=\left\{%
\begin{array}{cc}
-\mathrm{arcsin}(\frac{2 \Sigma_{\varphi,~12}}{\sqrt{\gamma}}), & \mbox{ if }
\, \Sigma_{\varphi,~11}<\Sigma_{\varphi,~22}, \\
\pi+\mathrm{arcsin}(\frac{2\Sigma_{\varphi,~12}}{\sqrt{\gamma}}), &
\mbox{
if } \, \Sigma_{\varphi,~11}>\Sigma_{\varphi,~22},%
\end{array}%
\right.  \label{squang}
\end{eqnarray}
with $\gamma=(\Sigma_{\varphi,~22}-\Sigma_{\varphi,~11})^2+(2\Sigma_{%
\varphi,~12})^2$. Here $\Sigma_{\varphi, ij}$ ($i,~j=1,~2$) is the element
of the covariant matrix ${\mbox{\boldmath{$\Sigma$}}}_{\varphi}$ as defined
in Eq.~(\ref{covarmat}). %An alternative quantity has been employed
%to measure the mechanical squeezing in Ref.~\cite{masj_aga14}.

%%%% Figure 6:
%%%%%%%%%%%%%%%%%%%%%%%%
\begin{figure}[!tb]
\centering
\includegraphics[width=2.5in]{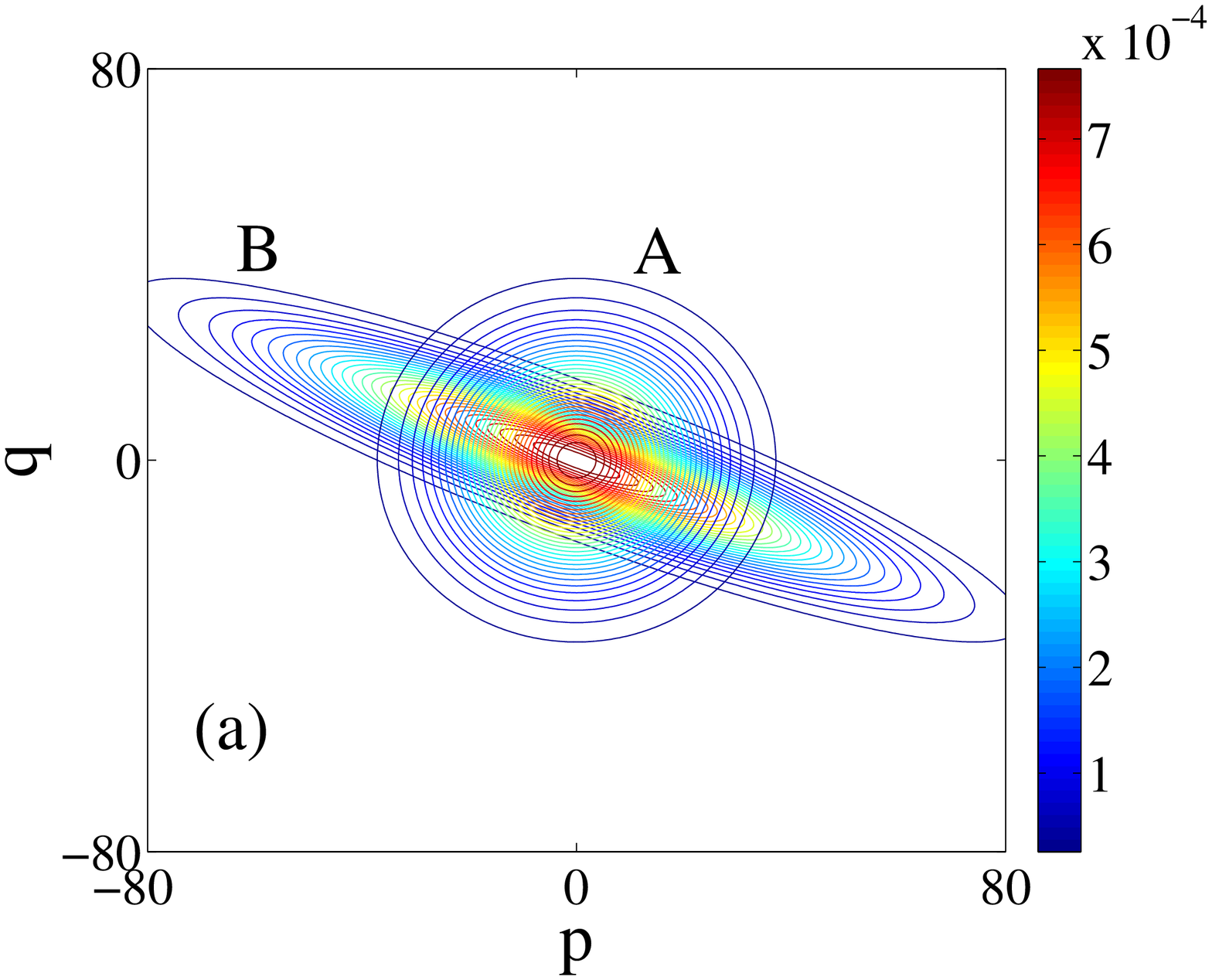} %
\includegraphics[width=2.5in]{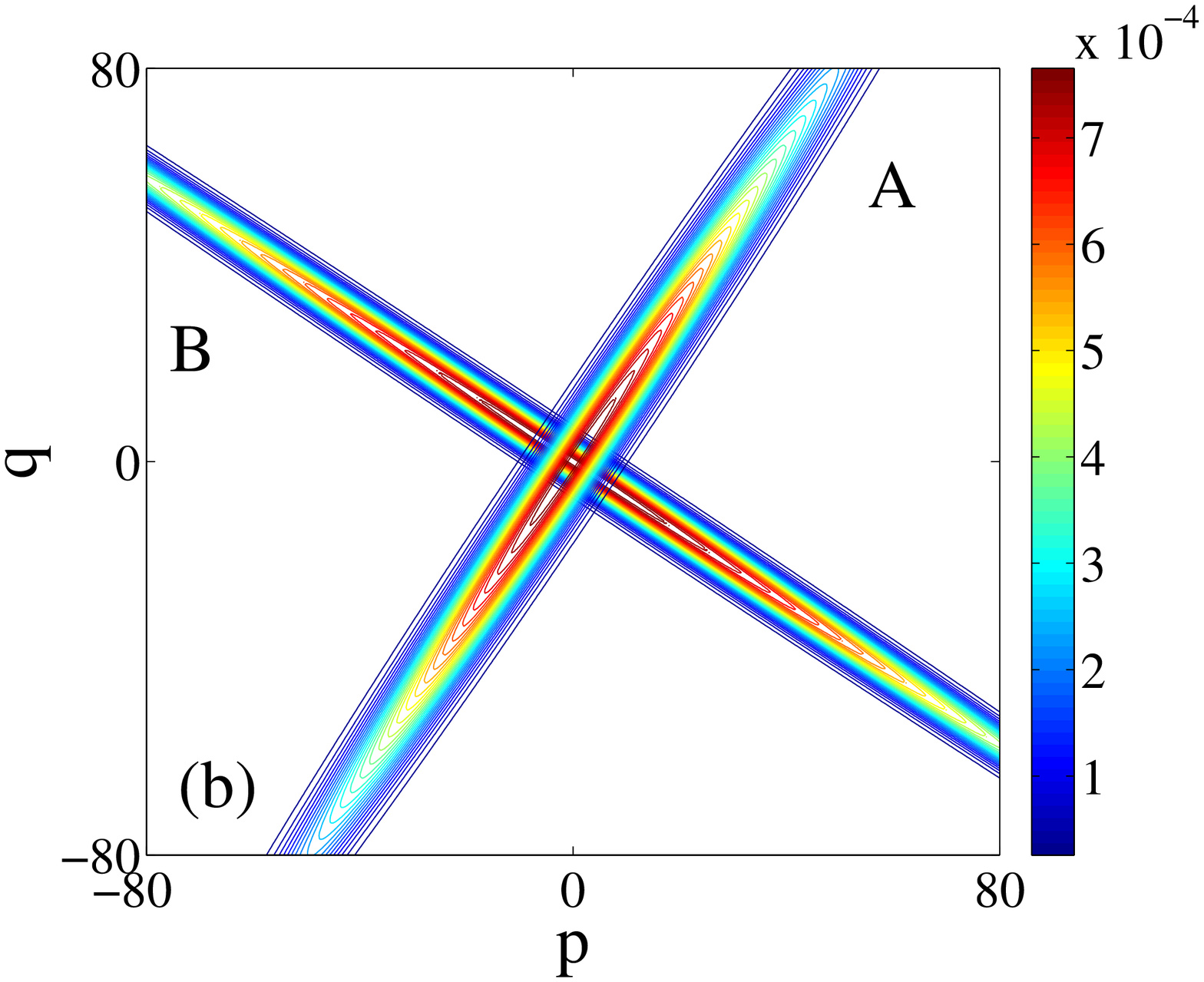} %
\includegraphics[width=2.5in]{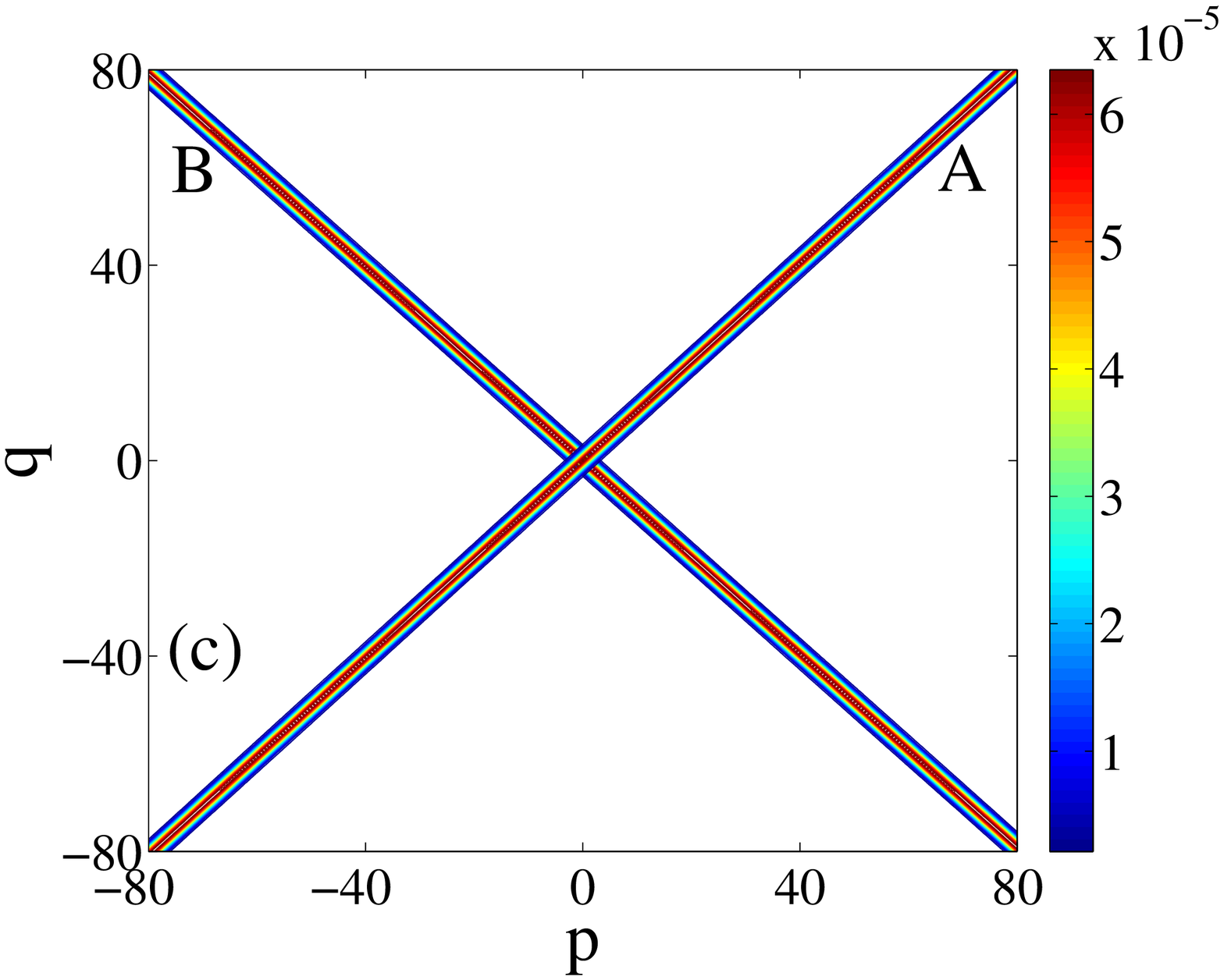} \hspace{3.5cm}
\caption{(Color online) The contour of Wigner functions of the mechanical
resonator before (after) the n-th kicked pulses are labeled as A (B). (a),
(b) and (c) corresponds to the pulse number $n=1$, $n=3$ and $n=10^3$,
respectively. It is obvious that here the mechanical squeezing can be
strengthened by the kicks. The parameters are $k=4$, $n_{th}=100$, and $%
\protect\theta=1.0$. }
\label{wigfunA}
\end{figure}
%%%%%%%%%%%%%%%%%%%%%%%%
%%%%%%%%%%%%%%%%%%%%%%%%

%%%% Figure 7:
%%%%%%%%%%%%%%%%%%%%%%%%
\begin{figure}[!tb]
\centering
\includegraphics[width=2.5in]{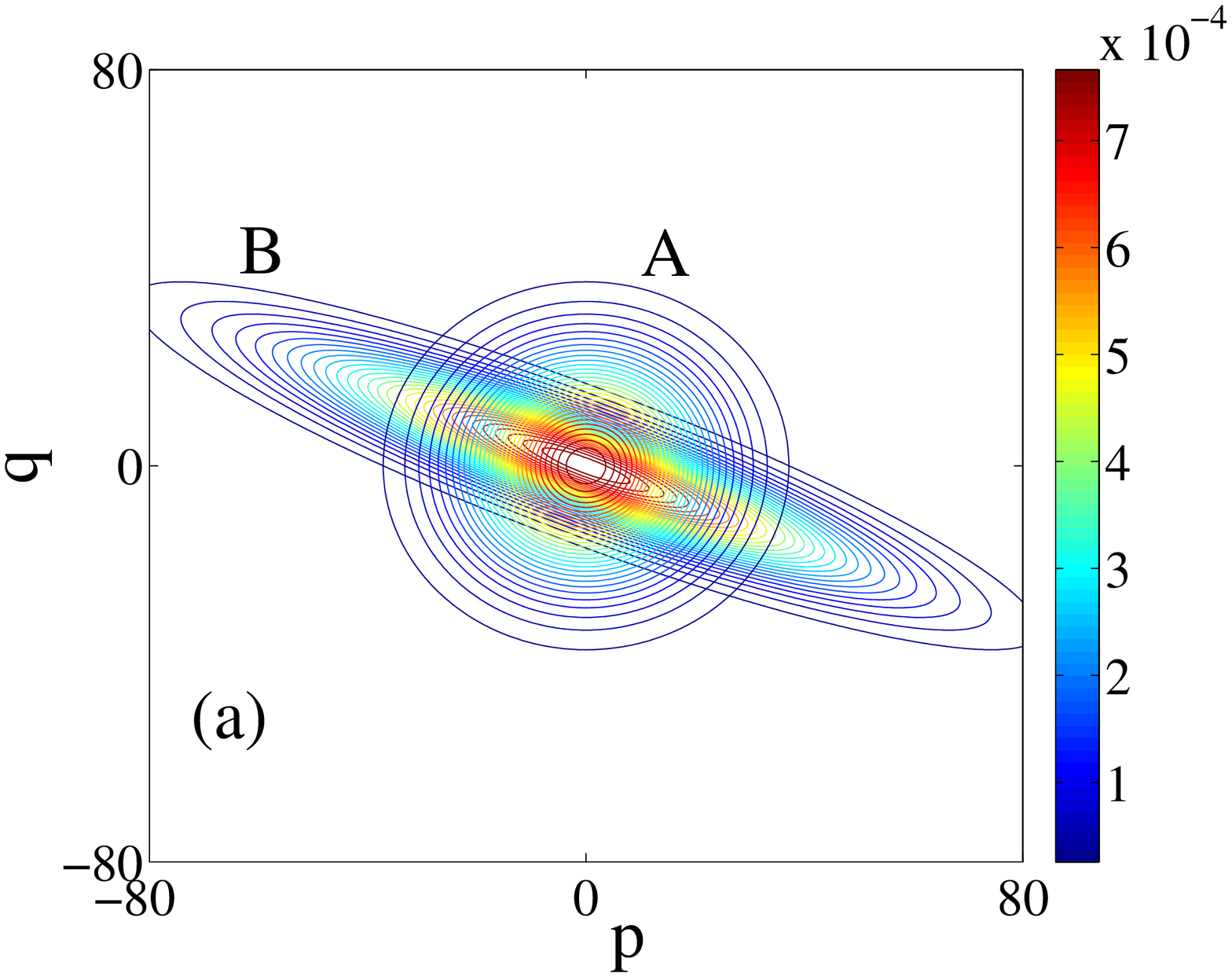} %
\includegraphics[width=2.5in]{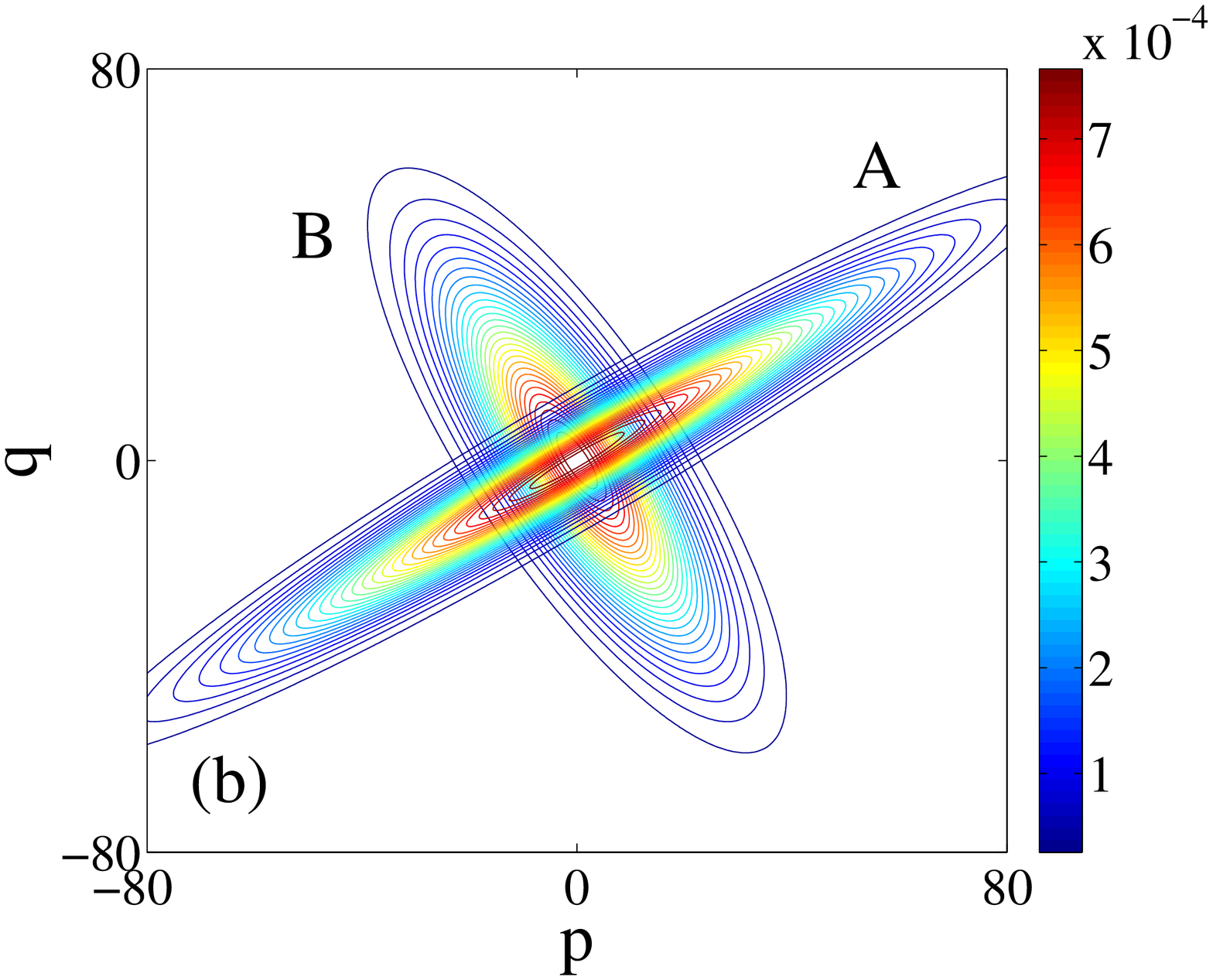} %
\includegraphics[width=2.5in]{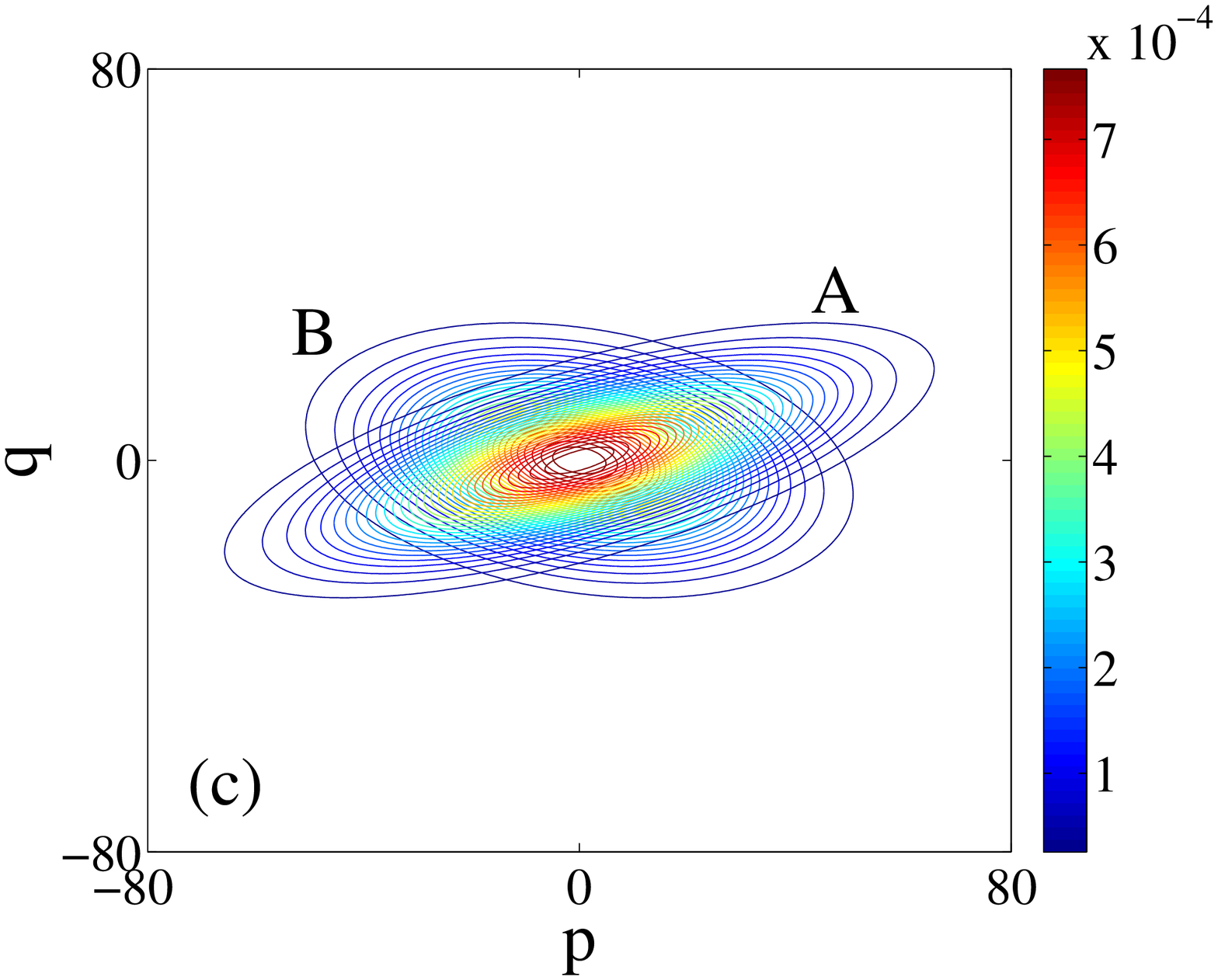} \hspace{3.5cm}
\caption{(Color online) The same to Fig.~\protect\ref{wigfunA} except for
the parameter $k=5$, corresponding to the free rotation angle $\protect%
\vartheta=2\protect\pi/5$. In this case, the kicks cannot effectively
produce the mechanical squeezing. }
\label{wigfunB}
\end{figure}
%%%%%%%%%%%%%%%%%%%%%%%%
%%%%%%%%%%%%%%%%%%%%%%%%

By rearranging the second-order moments given in Eq.~(\ref{solB}) as the
covariant matrix, the squeezing strength $r$ and the squeezing angle $\phi$
are obtained according to Eqs.~(\ref{squstreng}) and (\ref{squang}). Note
that $\alpha=0$ for the mechanical resonator studied in this paper. In Fig.~%
\ref{fishfigA}(b), we plot the mechanical squeezing strength $r$ as a
function of the pulse number $n$ for $k \leq 4$.
%\textbf{The lines of this figure are one-to-one correspondence to that of Fig.~\ref{fishfigA}(a).  -> }
By combining Fig.~\ref{fishfigA}(b) and Fig.~\ref{fishfigA}(a), it is
apparent that both the mechanical squeezing and the QFI are enhanced with
the increase of $k$. As a result, this squeezing as a quantum metrology
resource strengthens the QFI $F$. Although Fig.~\ref{fishfigA}(b) only shows
the case $k \leq 4$, the similar results have been found for the other $k$
values (not displayed here).

Based on the correlation between the QFI and the mechanical squeezing, we
provide an intuitive way to understand the QFI for $k=4$ (corresponding to a
free rotation time $\tau=T_{0}/4$) being optimal with the kick strength $%
\theta=1$. To this aim, it is useful to investigate the dynamics of the
mechanical Wigner function, obtained according to Eq.~(\ref{Wig_comat}).

The kick operator $\mathbf{K}$ produces the mechanical squeezing, which
remains invariant under free rotation $\mathbf{M}_{0}(\tau)$ (neglecting the
mechanical damping). After the first kick acts on the mechanical resonator,
the initial mechanical thermal state translates into a squeezed state with
the squeezing angle $\phi(n=1)\simeq \pi/8$, as shown in Fig.~\ref{wigfunA}%
(a). Here $\phi$ is defined as the angle between the $q$-axis and the
direction of the squeezed quadrature. Then, the squeezed state is rotated by
$\vartheta=\omega_{m} T_{0}/4=\pi/2$ along the clockwise direction by the
free evolution $\mathbf{M}_{0}(\tau=T_{0}/4)$. Under the effect of the
following kicks, the squeezing strength of the mechanical resonator
progressively increases, as displayed in Figs.~\ref{wigfunA}(b) and~\ref%
{wigfunA}(c).

Moreover, the squeezing angle is also gradually approaches to $\phi \simeq
\pi /4$ under some (e.g., $n\approx 10^{2}$) repetitive kicks (as well as
the free evolution between the kicks). Once the squeezing angle becomes to $%
\phi =\pi /4$, which coincides with the counterpart angle produced by the
squeezing action $\tilde{U}_{\theta =1}=\exp [-i(\hat{a}^{2}+h.c.)/2]$ in
the kick operator $U_{\text{K}}$, we find numerically that it will remain
unchanged for large $n$ limit. We would like to point out that here the
matching between the squeezing angle by kick and the free rotation angle
plays an essential role in strengthening the mechanical squeezing. As a
consequence, the QFI of the mechanical resonator is also enhanced.

We also show the mechanical Wigner function for the case $k=5$ in Fig.~\ref%
{wigfunB}, which corresponds to a free rotation angle $\vartheta=\omega_{m}
T_{0}/5=2\pi/5$. Thus, after the free rotation (with considering the
mechanical damping) represented by the operator $\mathbf{M}%
_{0}(\tau=T_{0}/5) $, the mechanical resonator cannot evolve into the
squeezed state with the squeezing angle $\phi=\pi/4$ by the kick operator $%
U_{\text{K}}$ in the long-time limit. This enables the kicks squeeze the
mechanical resonator ineffectively, as display in Fig.~\ref{wigfunB}(c).

\section{Conclusion}

In summary, the quantum pulsed optomechanical system is proposed to apply
for the quantum metrology in context of quantum parameter estimation by
focusing on investigating the Quantum Fisher information.
%is adopted to quantify the parameter-estimation precision.
We find that the mechanical frequency can be estimated with very high
precision if the mechanical period matches to that of the driving pulses. We
also display that the mechanical squeezing is the quantum resource used in
optimal quantum estimation on the frequency. In future, it is an interesting
subject to utilize coherence of the multi-mode cavity optomechanics~\cite%
{xwx13, FMass12} to enhance the accuracy of quantum parameter estimation.
\newline

\textit{Acknowledgments.} We thank X. G. Wang, X. W. Xu, H. Fu and X. Xiao for
their helpful discussions. This work is supported by the National Natural
Science Foundation of China (Grant Nos.~11365006, 11422437, and 11121403, 11565010)
and the 973 program (Grants No. 2012CB922104 and No. 2014CB921403),
Guizhou province science and technology innovation talent team (Grant No. (2015)4015).\newline

%\newpage
\appendix

\section{QFI of a single-mode Gaussian state}

In order to keep the completeness of this paper, here we review the main
aspects of local quantum estimation theory, especially focusing on the QFI
of a single-mode Gaussian state. Note that the mechanical resonator studied
in this paper stays in a single-mode Gaussian state.

Let $\varphi$ denote a single parameter to be estimated, and $%
p(\zeta|\varphi)$ be the probability density with measurement outcome $\{
\zeta \}$ for a continuous observable $W$ conditioned on the fixed parameter
$\varphi$. The value of the parameter $\varphi$ can be inferred from the estimator
function $\hat{\varphi}=\hat{\varphi}(\zeta_{1},~...,~\zeta_{N})$, based on the measurement results
$\zeta_{1}$,~...,~$\zeta_{N}$ of $N$ replicas of the system. 
Usually, this is achieved by the maximum likelihood estimation.
With the definition of the classical Fisher information \cite{Fisher}
\begin{equation}
\begin{array}{llll}
H_{\varphi}= \int d \zeta p(\zeta|\varphi)[\frac{\partial}{\partial \varphi}
\ln p(\zeta|\varphi)]^2, &  &  &
\end{array}%
\end{equation}
the classical Cram\'{e}r-Rao inequality~\cite{Holevo} gives the bound of the
variance $\mathrm{Var}(\hat{\varphi})$ for an unbiased estimator $\hat{%
\varphi}$
\begin{equation}
\begin{array}{llll}
\mathrm{Var}(\hat{\varphi}) \geq \frac{1}{H_{\varphi}}, &  &  &
\end{array}%
\end{equation}

Extending from classical to quantum regime, the conditional probability $%
p(\zeta|\varphi)$ is determined by positive operator valued measure operator
$\{ \hat{E}_{\zeta} \}$ for a parameterized quantum state $\rho_{\varphi}$, $%
p(\zeta|\varphi)=\mathrm{Tr}[\hat{E}_{\zeta} \rho_{\varphi}].$ To determine
the ultimate bound to precision posed by quantum mechanics, the Fisher
information must be maximized over all possible measurements~\cite{paris09}.

By introducing the symmetric logarithmic derivative $L_{\varphi}$ determined
by
\[
\frac{ \partial \rho_{\varphi} }{ \partial \varphi }=\frac{1}{2}
(\rho_{\varphi} L_{\varphi}+ L_{\varphi} \rho_{\varphi} ),
\]
the so-called quantum Cram\'{e}r-Rao inequality gives a bound to the
variance of any unbiased estimator~\cite{Caves94}:
\begin{equation}
\begin{array}{llll}
\mathrm{Var}(\hat{\varphi}) \geq \frac{1}{H_{\varphi}} \geq \frac{1}{%
F_{\varphi}}, &  &  &
\end{array}
\label{crineqB}
\end{equation}
where
\begin{equation}
\begin{array}{llll}
F_{\varphi}=\mathrm{Tr}[\rho_{\varphi} L_{\varphi}^2] &  &  &
\end{array}%
\end{equation}
is the quantum Fisher information.

The two bounds for the precision of parameter estimation \cite{Caves96} have
been found, the so-called shot-noise limit $1/\sqrt{N}$ and the Heisenberg
limit 1/$N$. Usually, here $N$ is the total particle number contributed to
quantum estimation.

It is not easy to give the explicit formula of QFI for a general system.
Fortunately, the QFI is related to the Bures distance~\cite{Caves94} through
\begin{equation}
\begin{array}{llll}
D_{B}^2[\rho_{\varphi}, \rho_{\varphi+d\varphi}]= \frac{1}{4}F_{\varphi}d
\varphi^2, &  &  &
\end{array}
\label{fdfisher}
\end{equation}
where the definition of the Bures distance between two quantum states $\rho$
and $\sigma$ is as~\cite{Nielsen00}
\[
D_{B}[\rho, \sigma]=[ 2(1-\mathrm{Tr}\sqrt{\rho^{1/2} \sigma \rho^{1/2}}%
)]^{1/2}.
\]

The Wigner function for an arbitrary given state $\rho$, defined as
\[
W(q, p)=\frac{1}{2 \pi}\int_{-\infty}^{\infty}~ds e^{-ips}\langle
q-s|\rho|q+s\rangle,
\]
can be used to equivalently represent the corresponding quantum state $\rho$%
, as there is a one-to-one correspondence between them.
%The Wigner function
%has the two appealing properties: (1) its marginal distributions are the true probability distributions; and
%(ii) the overlap between two states equals the integral of the products of their Wigner functions.
%Many (experimental) methods have been proposed to measure or reconstruct the Wigner function of the state %because it plays a key role in various field, such as the quantum-classical transition, non-classical of % quantum state, quantum information processing, etc.
Gaussian state, as a specific kind of continuous-variable states, has wide
applications in actual quantum information processing \cite{mgap05}. They
can be reproduced efficiently and unconditionally in the experiment. The
unconditionedness is one advantage of the continuous-variable state, which
is hard to achieve in the qubit-based discrete-variable.

A state is said to be Gaussian in case its Wigner function is Gaussian. A
Gaussian state can be completely characterized by the first-order moment and
the second-order moment
\begin{equation}
\begin{array}{llll}
\overline{X_{i}}= \langle \hat{X}_{i}\rangle,
\\
\\
\Sigma_{\varphi,~ij}=\frac{1}{2}\langle (\hat{X}_{i} \hat{X}_{j}+\hat{X}_{j}
\hat{X}_{i}) \rangle- \langle \hat{X}_{i} \rangle \langle \hat{X}_{j} \rangle
&  &  &  \\
~~~~~~~~~~=\int W(\vec{X}){X}_{i}{X}_{j} d^2 \vec{X}. &  &  &
\end{array}
\label{covarmat}
\end{equation}
Here
\begin{equation}
\begin{array}{llll}
\vec{\hat{X}} \equiv (\hat{q}, \hat{p}),~~~ \langle \cdots \rangle \equiv
\mathrm{Tr}(\rho_{\varphi} \cdots), &  &  &
\end{array}%
\end{equation}
and $\varphi$ is the parameter to be estimated in the quantum state $%
\rho_{\varphi}$. The Wigner function is related to the the second-order
moments as (setting the first-order moments being zeros)
\begin{equation}
\begin{array}{llll}
W(\vec{X})= \frac{1}{ 2 \pi \sqrt{ \mathrm{Det} {\mbox{\boldmath{$\Sigma$}}%
_{\varphi}} } } \exp( -\frac{1}{2}\vec{X} {\mbox{\boldmath{$\Sigma$}}%
_{\varphi}}^{-1} \vec{X}^{T}). &  &  &
\end{array}
\label{Wig_comat}
\end{equation}
\newline
Based on the fidelity between arbitrary single-mode Gaussian states $%
\rho_{1} $ and $\rho_{2}$,
\begin{equation}
\begin{array}{llll}
f(\rho_{1}, \rho_{2})= \frac{2 \exp[-\frac{1}{2} \Delta X^{T} ({%
\mbox{\boldmath{$\Sigma$}}}_{1}+{\mbox{\boldmath{$\Sigma$}}}_{2} )^{-1}
\Delta X ]}{\sqrt{| {\mbox{\boldmath{$\Sigma$}}}_{1}+{\mbox{\boldmath{$%
\Sigma$}}}_{2}|+(1-|{\mbox{\boldmath{$\Sigma$}}}_{1}|)(1-|{%
\mbox{\boldmath{$\Sigma$}}}_{2}|)} -\sqrt{(1-|{\mbox{\boldmath{$\Sigma$}}}%
_{1}|)(1-|{\mbox{\boldmath{$\Sigma$}}}_{2}|)}}, &  &  &
\end{array}
\label{fdgau}
\end{equation}
making use of Eq.~(\ref{fdfisher}), the QFI of the single-mode Gaussian
state is found to be \cite{opin13, dddsouz14}
\begin{equation}
\begin{array}{llll}
F_{\varphi}=\frac{\mathrm{Tr}[( {\mbox{\boldmath{$\Sigma$}}}_{\varphi}^{-1}{%
\mbox{\boldmath{$\Sigma$}}}_{\varphi}^{\prime 2}]}{2(1+P_{\varphi}^2)}+ 2%
\frac{P_{\varphi}^{\prime 2}}{1-P_{\varphi}^4 }+ \Delta \vec{X}%
_{\varphi}^{\prime ~T}{\mbox{\boldmath{$\Sigma$}}}_{\varphi}^{-1} \Delta
\vec{X}_{\varphi}^{\prime }. &  &  &
\end{array}
\label{fisherGau}
\end{equation}
Here $\Delta \vec{X}= \langle \vec{X}_{1}-\vec{X}_{2} \rangle $ is the mean
relative displacement, $P_{\varphi}=|{\mbox{\boldmath{$\Sigma$}}}%
_{\varphi}|^{-1/2}$ denotes the purity of the state, and
\begin{equation}
\begin{array}{llll}
\Delta \vec{X}_{\varphi}^{\prime }= d \langle \vec{X}_{\varphi+\epsilon} -%
\vec{X}_{\varphi}\rangle/ d \epsilon |_{\epsilon=0}. &  &  &
\end{array}
\label{fsdiff}
\end{equation}
We notice that the quantum Cram$\acute{e}$r-Rao bound of two-mode Gaussian
states \cite{mggen08} was investigated previously.

%%%%%%%%%%%%%%%%%%%%%%%%%%
%\newpage

%\end{CJK*}


\begin{thebibliography}{999}
\bibitem{VGio11} V. Giovanetti, S. Lloyd and L. Maccone, Science \textbf{306}%
, 1330 (2004); V. Giovannetti, S. Lloyd and L. Maccone, Phys. Rev. Lett.
\textbf{96}, 010401 (2006); V. Giovannetti, S. Lloyd, and L. Maccone, Nat.
Photonics \textbf{5}, 222 (2011).

\bibitem{Smerzi09} L. Pezz$\acute{e}$ and A. Smerzi, Phys. Rev. Lett.
\textbf{102}, 100401 (2009); A. Smerzi, Phys. Rev. Lett. \textbf{109},
150410 (2012).

\bibitem{xmLu10} X. M. Lu, X. G. Wang, and C. P. Sun, Phys. Rev. A \textbf{82%
}, 042103 (2010); W. Zhong, Z. Sun, J. Ma, X. G. Wang, and F. Nori, Phys.
Rev. A \textbf{87}, 022337 (2013); Q. S. Tan, Y. X. Huang, X. L. Yin, L. M.
Kuang, and X. G. Wang, Phys. Rev. A \textbf{87}, 032102 (2013).

\bibitem{yy14b} Y. Yao, L. Ge, X. Xiao, X. G. Wang, and C. P. Sun, Phys.
Rev. A 90, 022327 (2014).

\bibitem{jin13} Y. M. Zhang, X. W. Li, W. Yang, and G. R. Jin, Phys. Rev. A
\textbf{88}, 043832 (2013).

\bibitem{qz15} Q. Zheng, L. Ge, Y. Yao, and Q. J. Zhi, Phys. Rev. A \textbf{%
91}, 033805 (2015).

% optical system

\bibitem{Nagata07} T. Nagata, R. Okamoto, J. L. OBrien, K. Sasaki, and S.
Takeuchi, Science \textbf{316}, 726 (2007).

\bibitem{Afek10} I. Afek, O. Ambar, and Y. Silberberg, Science \textbf{328},
879 (2010).

\bibitem{zyou14} F. Hudelist, J. Kong, C. Liu, J. Jing, Z. Y. Ou, and W. P.
Zhang, Nat. Commun. \textbf{5}, 3049 (2014).

%BEC

\bibitem{Strobel14} H. Strobel, W. Muessel, D. Linnemann, T. Zibold, D. B.
Hume, L. Pezz\`{e}, A. Smerzi, and M. K. Oberthaler, Science \textbf{345},
424 (2014).

\bibitem{wdli14} W. D. Li, T. C. He, and A. Smerzi, Phys. Rev. Lett. \textbf{%
113}, 023003 (2014).

%NV

\bibitem{XYpan15} G. Q. Liu, Y. R. Zhang, Y. C. Chang, J. D. Yue, H. Fan, X.
Y. Pan, Nat. Commun. \textbf{6}, 6726 (2015).

\bibitem{zhao13} N. Zhao and Z. Q. Yin, Phys. Rev. A \textbf{90}, 042118
(2014).

% opto_meto

\bibitem{kiwa13} K. Iwasawa, K. Makino, H. Yonezawa, M. Tsang, A. Davidovic,
E. Huntington, and A. Furusawa, Phys. Rev. Lett. \textbf{111}, 163602 (2013).

\bibitem{szang13} S. Z. Ang, G. I. Harris, W. P. Bowen, and M. Tsang, New.
J. Phys. \textbf{15}, 103028 (2013).

\bibitem{mtsan13} M. Tsang, New. J. Phys. \textbf{15}, 073005 (2013).

% optomech

\bibitem{tjkip08} T. J. Kippenberg and K. J. Vahala, Science \textbf{321},
1172 (2008).

\bibitem{pmey13} M. Aspelmeyer, P. Meystre, and K. C. Schwab, Phys. Today,
\textbf{65}, 29 (2012); P. Meystre, Ann. Phys. (Berlin) \textbf{525},
215(2013).

\bibitem{matjkfm14} M. Aspelmeyer, T. J. Kippenberg, and F. Marquardt,
\textit{Cavity Optomechanics} (Springer-Verlag Berlin Heidelberg, 2014).

\bibitem{masp14} M. Aspelmeyer, T. J. Kippenberg, and F. Marquardt, Rev.
Mod. Phys. \textbf{86}, 1391 (2014).

% optomech: macroscopic tunneling

\bibitem{lfbuch13} L. F. Buchmann, L. Zhang, A. Chiruvelli, and P. Meystre
Phys. Rev. Lett. \textbf{108}, 210403 (2012); H. T. Tan, F. Bariani, G. X.
Li, and P. Meystre, Phys. Rev. A \textbf{88}, 023817 (2013).

\bibitem{wmar03} W. Marshall, C. Simon, R. Penrose, and D. Bouwmeester,
Phys. Rev. Lett. \textbf{91}, 130401 (2003).

\bibitem{cpsun07} F. Xue, Y. X. Liu, C. P. Sun, and F. Nori, Phys. Rev. B
\textbf{76}, 064305 (2007).

% precise measure

\bibitem{DRug04} D. Rugar, R. Budakian, H. J. Mamin, and B. W. Chui, Nature
(London) \textbf{430}, 329 (2004); A. G. Krause, M. Winger, T. D. Blasius,
Q. Lin, and O. Painter, Nat. Photon. \textbf{6} 768 (2012).

\bibitem{careg08} C. A. Regal, J. D. Teufel, and K. W. Lehnert, Nat. Phys.
\textbf{4}, 555 (2008).

\bibitem{ganet09} J. D. Teufel, T. Donner, M. A. Castellanos-Beltran, J. W.
Harlow, and K. W. Lehnert, Nat. Nanotechnol. \textbf{4}, 820 (2009).

\bibitem{sfors12} S. Forstner, S. Prams, J. Knittel, E. D. van Ooijen, J. D.
Swaim, G. I. Harris, A. Szorkovszky, W. P. Bowen, and H. Rubinsztein-Dunlop,
Phys. Rev. Lett. \textbf{108}, 120801 (2012).

\bibitem{jmtay14} X. Xu and J. M. Taylor, Phys. Rev. A \textbf{90}, 043848
(2014).

\bibitem{aarv13} A. Arvanitaki and A. A. Geraci, Phys. Rev. Lett. \textbf{110%
}, 071105 (2013).

\bibitem{sman03} S. Mancini, D. Vitali, and P. Tombesi, Phys. Rev. Lett.
\textbf{90}, 137901 (2003).

% quant illum

\bibitem{shbar15} Sh. Barzanjeh, S. Guha, C. Weedbrook, D. Vitali, J. H.
Shapiro, and S. Pirandola, Phys. Rev. Lett. \textbf{114}, 080503 (2015).

% EIT

\bibitem{sweis10} S. Weis, R. Rivi\`{e}re, S. Del\'{e}glise, E. Gavartin, O.
Arcizet, A. Schliesser, and T. J. Kippenberg, Science \textbf{330}, 1520
(2010); A. H. Safavi-Naeini, T. P. M. Alegre, J. Chan, M. Eichenfield, M.
Winger, Q. Lin, J. T. Hill, D. E. Chang, and O. Painter, Nature (London)
\textbf{472}, 69 (2011).

\bibitem{gsash10} G. S. Agarwal and S. Huang, Phys. Rev. A \textbf{81},
041803(R) (2010).

% cooling

\bibitem{iwil07} I. Wilson-Rae, N. Nooshi, W. Zwerger, and T. J. Kippenberg,
Phys. Rev. Lett. \textbf{99}, 093901 (2007); F. Marquardt, J. P. Chen, A. A.
Clerk, and S. M. Girvin, ibid. \textbf{99}, 093902 (2007).

\bibitem{cgenes08} C. Genes, D. Vitali, P. Tombesi, S. Gigan, and M.
Aspelmeyer, Phys. Rev. A \textbf{77}, 033804 (2008).

\bibitem{jdteufel10} J. D. Teufel, T. Donner, D. Li, J. H. Harlow, M. S.
Allman, K. Cicak, A. J. Sirois, J. D. Whittaker, K. W. Lehnert, and R. W.
Simmonds, Nature (London) \textbf{475}, 359 (2011).

\bibitem{yli14} Y. J. Guo, K. Li, W. J. Nie, and Y. Li, Phys. Rev. A \textbf{%
90}, 053841 (2014).

\bibitem{jctpm11} J. Chan, T. P. Mayer Alegre, A. H. Safavi-Naeini, J. T.
Hill, A. Krause, S. Groblacher, M. Aspelmeyer and O. Painter, Nature \textbf{%
478}, 89 (2011).

% entangle

\bibitem{mpater07} M. Paternostro, D. Vitali, S. Gigan, M. S. Kim, C.
Brukner, J. Eisert, and M. Aspelmeyer, Phys. Rev. Lett. \textbf{99}, 250401
(2007); Sh. Barzanjeh, M. Abdi, G. J. Milburn, P. Tombesi, and D. Vitali,
Phys. Rev. Lett. \textbf{109}, 130503 (2012).

\bibitem{ltydw13} L. Tian, Phys. Rev. Lett. \textbf{110}, 233602 (2013); Y.
D. Wang and A. A. Clerk, ibid. \textbf{110}, 253601 (2013).

\bibitem{wwiesghof15} W. Wieczorek, S. G. Hofer, J. Hoelscher-Obermaier, R.
Riedinger, K. Hammerer, and M. Aspelmeyer, Phys. Rev. Lett. \textbf{114},
223601 (2015).

% pulse optomech

\bibitem{vanner11} M. R. Vanner, I. Pikovski, G. D. Cole, M. S. Kim, \v{C}.
Bruknera, K. Hammerer, G. J. Milburn, and M. Aspelmeyer, Proc. Natl. Acad.
Sci. USA \textbf{108}, 16182 (2011).

\bibitem{sl98} L. Viola and S. Lloyd, Phys. Rev. A 58, 2733 (1998).

\bibitem{khammer05} K. Hammerer, E. S. Polzik, and J. I. Cirac, Phys. Rev. A
\textbf{72}, 052313 (2005).

\bibitem{oromeris11} O. Romero-Isart, A. C. Pflanzer, M. L. Juan, R.
Quidant, N. Kiesel, M. Aspelmeyer, and J. I. Cirac, Phys. Rev. A \textbf{83}%
, 013803 (2011).

\bibitem{masj_aga14} M. Asjad, G. S. Agarwal, M. S. Kim, P. Tombesi, G. Di
Giuseppe, and D. Vitali, Phys. Rev. A \textbf{89}, 023849 (2014).

\bibitem{jqliao11} J. Q. Liao and C. K. Law, Phys. Rev. A \textbf{84},
053838 (2011).

\bibitem{hofer11} S. G. Hofer, W. Wieczorek, M. Aspelmeyer, and K. Hammerer,
Phys. Rev. A \textbf{84}, 052327 (2011).

\bibitem{tapal13} T. A. Palomaki, J. D. Teufel, R. W. Simmonds, K. W.
Lehnert, Science 342, 710 (2013).

\bibitem{qyhe13} Q. Y. He, and M. D. Reid, Phys. Rev. A \textbf{88}, 052121
(2013); S. Kiesewetter, Q. Y. He, P. D. Drummond, and M. D. Reid, Phys. Rev.
A \textbf{90}, 043805 (2014).

\bibitem{vanner13} M. R. Vanner, J. Hofer, G. D. Cole, and M. Aspelmeyer,
Nat. Commun \textbf{4}, 2295 (2013).

\bibitem{smach12} S. Machnes, J. Cerrillo, M. Aspelmeyer, W. Wieczorek, M.
B. Plenio, and A. Retzker, Phys. Rev. Lett. \textbf{108}, 153601 (2012).

% phys implem

\bibitem{oarc06} O. Arcizet, P. F. Cohadon, T. Briant, M. Pinard, A.
Heidmann, Nature (London) \textbf{444}, 71 (2006).

\bibitem{gantk09} G. Anetsberger, O. Arcizet, Q. P. Unterreithmeier, R. Rivi%
\`{e}re, A. Schliesser, E. M. Weig, J. P. Kotthaus, and T. J. Kippenberg,
Nat. Phys. \textbf{5}, 909 (2009).

\bibitem{kies13} N. Kiesel, F. Blaser, U. Deli$\acute{c}$, D. Grass, R.
Kaltenbaek, and M. Aspelmeyer, Proc. Natl. Acad. Sci. USA \textbf{110},
14180 (2013).

\bibitem{tppurdy10} T. P. Purdy, D. W. C. Brooks, T. Botter, N. Brahms,
Z.-Y. Ma, and D. M. Stamper-Kurn, Phys. Rev. Lett. \textbf{105}, 133602
(2010).

\bibitem{jcsan10} J. C. Sankey, C. Yang, B. M. Zwickl, A. M. Jayich, and J.
G. E. Harris, Nat. Phys. \textbf{6}, 707 (2010).

\bibitem{jdthom08} J. D. Thompson, B. M. Zwickl, A. M. Jayich, F. Marquardt,
S. M. Girvin, J. G. E. Harris, Nature (London) \textbf{452}, 72 (2008).

\bibitem{dengli08} Z. J. Deng, Y. Li, M. Gao, and C. W. Wu, Phys. Rev. A
\textbf{78}, 032303 (2008).

\bibitem{ghein10} G. Heinrich, J. G. E. Harris, F. Marquardt, Phys. Rev. A
\textbf{81}, 011801 (2010); H. Z. Wu, G. Heinrich, and F. Marquardt, New J.
Phys. \textbf{15}, 123022(2013).

% optomech crystral
\bibitem{tkpmk15} T. K. Paraiso, M. Kalaee, L. Zang, H. Pfeifer, F.
Marquardt, and O. Painter, Phys. Rev. X \textbf{5}, 041024 (2015).



\bibitem{pmeys08} M. Bhattacharya and P. Meystre, Phys. Rev. Lett. \textbf{99%
}, 073601 (2007); M. Bhattacharya, H. Uys, and P. Meystre, Phys. Rev. A
\textbf{77}, 033819 (2008).

\bibitem{ankb08} A. Nunnenkamp, K. B{\o }rkje, J. G. E. Harris, and S. M.
Girvin, Phys. Rev. A \textbf{82}, 021806(R) (2010).

%mass spect

\bibitem{kdzhu13} J. J. Li and K. D. Zhu, Phys. Rep. \textbf{525}, 223
(2013).

%parameter

\bibitem{neflow12} N. E. Flowers-Jacobs, S.W. Hoch, J. C. Sankey, A.
Kashkanova, A. M. Jayich, C. Deutsch, J. Reichel, and J. G. E. Harris, Appl.
Phys. Lett. \textbf{101}, 221109 (2012).

% pulse QFI

\bibitem{Ueda10} Y. Watanabe, T. Sagawa, and M. Ueda, Phys. Rev. Lett.
\textbf{104}, 020401 (2010).

\bibitem{tan13} Q. S. Tan, Y. X. Huang, X. L. Yin, L. M. Kuang, and X. G.
Wang, Phys. Rev. A \textbf{87}, 032102 (2013).

\bibitem{sbad082} S. Boixo, A. Datta, S. T. Flammia, A. Shaji, E. Bagan, and
C. M. Caves, Phys. Rev. A \textbf{77}, 012317 (2008); Y. C. Liu, G. R. Jin,
and L. You, Phys. Rev. A \textbf{82}, 045601 (2010).

% experiment proposal




\bibitem{garpol00} C. W. Gardiner and P. Zoller, \textit{Quantum Noise}
(Springer, New York, 2000).

% squeeze metrolog

\bibitem{jmaxgw11} J. Ma, X. G. Wang, C. P. Sun, and F. Nori, Phys. Rep.
\textbf{509}, 89 (2011).

\bibitem{caves81} C. M. Caves, Phys. Rev. D 23, 1693 (1981); M. D. Lang and
C. M. Caves, Phys. Rev. Lett. 111, 173601 (2013).

\bibitem{asmer13} L. Pezz\'{e} and A. Smerzi, Phys. Rev. Lett. 110, 163604
(2013).

\bibitem{cweed12} C. Weedbrook, S. Pirandola, R. Garcia-Patron, N. J. Cerf,
T. C. Ralph, J. H. Shapiro, and S. Lloyd, Rev. Mod. Phys. \textbf{84}, 621
(2012).

\bibitem{xwx13} X. W. Xu, Y. J. Zhao, and Y. X. Liu, Phys. Rev. A \textbf{88}%
, 022325 (2013).



\bibitem{FMass12} F. Massel, S. U. Cho, J. M. Pirkkalainen, P. J. Hakonen,
T. T. Heikkil$\ddot{a}$, and M. Sillanp$\ddot{a}\ddot{a}$, Nat. Commun.
\textbf{3}, 987 (2012).

% fisher information

\bibitem{Fisher} R. A. Fisher, Proc. Cambridge Philos. Soc. \textbf{22}, 700
(1925).

\bibitem{Holevo} A. S. Holevo, \textit{Probabilistic and Statistical Aspects
of Quantum Theory} (North-Holland, Amsterdam, 1982).

\bibitem{paris09} M. G. A. Paris, Int. J. Quant. Inf. \textbf{7}, 125 (2009).

\bibitem{Caves94} S. L. Braunstein and C. M. Caves, Phys. Rev. Lett. \textbf{%
72}, 3439 (1994).

\bibitem{Caves96} S. L. Braunstein, C. M. Caves, and G. J. Milburn, Ann.
Phys. (N.Y.) \textbf{247}, 135 (1996).

\bibitem{Nielsen00} M. A. Nielsen and I. L. Chuang, \textit{Quantum
Computation and Quantum Information} (Cambridge University Press, Cambridge,
U.K., 2000).

%gaussian state

\bibitem{mgap05} A. Ferraro, S. Olivares, and M. G. A. Paris, \textit{%
Gaussian States in Quantum Information}, Napoli Series on Physics and
Astrophysics (Bibliopolis, Napoli, 2005).

% QFI gaussian

\bibitem{opin13} O. Pinel, P. Jian, N. Treps, C. Fabre, and D. Braun, Phys.
Rev. A \textbf{88}, 040102 (2013).

\bibitem{dddsouz14} D. D. de Souza, M. G. Genoni, M. S. Kim, Phys. Rev. A
\textbf{90}, 042119(2014).

% two mode gaussian state QFI

\bibitem{mggen08} M. G. Genoni, P. Giorda, and M. G. A. Paris, Phys. Rev. A
\textbf{78}, 032303 (2008); X. X. Zhang, Y. X. Yang, and X. B. Wang, Phys.
Rev. A \textbf{88}, 013838 (2013).
\end{thebibliography}
\end{document}